\documentclass[10pt,twocolumn]{article}
\usepackage[letterpaper,top=0.58in,bottom=0.60in,left=0.62in,right=0.62in,columnsep=0.27in]{geometry}
\usepackage{fontspec}
\usepackage{amsmath}
\usepackage{unicode-math}
\usepackage{microtype,graphicx,booktabs,array,calc,caption}
\usepackage{tabularray}
\usepackage{stfloats,balance,needspace,titlesec,enumitem}
\usepackage[normalem]{ulem}
\usepackage{xurl,hyperref}
\hypersetup{hidelinks,pdftitle={LMP-GNN: probabilistic reconstruction of missing lane counts for Signed Max-Pressure traffic signal control},pdfauthor={Zhihao Wan, Xiangle Pan, Xinqiang Chen, Gen Li, Qiang Luo}}
\newcommand{\ul}[1]{\uline{#1}}
\newcommand{\TableBodyFont}{\fontsize{9}{10.6}\selectfont}
\titleformat{\section}{\normalfont\bfseries\fontsize{11}{12.5}\selectfont}{}{0pt}{}
\titleformat{\subsection}{\normalfont\bfseries\itshape\fontsize{10}{11.8}\selectfont}{}{0pt}{}
\titleformat{\subsubsection}{\normalfont\bfseries\itshape\fontsize{10}{11.8}\selectfont}{}{0pt}{}
\titlespacing*{\section}{0pt}{8pt}{4pt}
\titlespacing*{\subsection}{0pt}{7pt}{3pt}
\titlespacing*{\subsubsection}{0pt}{6pt}{3pt}
\begin{document}

\twocolumn[\begin{@twocolumnfalse}

\begin{center}\vspace{-6pt}{\fontsize{16}{18.5}\selectfont\bfseries LMP-GNN: probabilistic reconstruction of missing lane counts for Signed Max-Pressure traffic signal control\par}
\vspace{3pt}
{\fontsize{10.5}{12.6}\selectfont Zhihao Wan, Xiangle Pan, Xinqiang Chen, Gen Li, Qiang Luo\par}\end{center}

\vspace{-4.6pt}

{\fontsize{9.5}{11.4}\selectfont Abstract: Missing lane-count observations can distort pressure-based signal decisions even when neighboring detectors remain operational. We propose a lane-movement probabilistic graph neural network (LMP-GNN) that uses the movement relations involved in pressure computation to predict a mean and standard deviation for each lane. Three rules convert these outputs into replacement counts using the mean alone, a fixed uncertainty discount, or a staleness-dependent discount. Observed counts remain unchanged, and the completed state is supplied to an unchanged Signed Max-Pressure controller. Evaluation covers reconstruction and uncertainty calibration, decision-time diagnostics, and closed-loop traffic performance. Across 4,333,392 masked lane events, reconstruction achieved a mean absolute error of 0.7873 vehicles per lane. In a separate stored-trace audit of 372 network--outage--seed cells, pressure-score error was strongly associated with phase disagreement, with a Spearman correlation of 0.929, identifying pressure fidelity as a key decision-level diagnostic. Across five fixed-demand CityFlow networks, the fixed-discount rule reduced accrued average travel time by up to 13.74\% relative to road-level mean imputation under correlated missingness. It also reduced travel time at 60\% random missingness, whereas mean imputation performed better at 80\% and 90\%. The selected model has 63,362 parameters and a median single-thread inference time of 0.983 ms on a central processing unit. These results support lightweight probabilistic lane reconstruction as a practical input to pressure-based control, with traffic benefits that depend on the missingness regime.\par}

\vspace{4pt}

{\fontsize{9.5}{11.4}\selectfont Index Terms---Graph neural networks, optimization and control, system state estimation, traffic networks, traffic signal control.\par}

\vspace{10pt}\end{@twocolumnfalse}]

\section{1. Introduction}\label{introduction}

Traffic signals allocate limited green time among competing movements at urban intersections. A traffic-responsive controller selects signal phases from current measurements rather than following only a preset timing plan. This responsiveness depends on reliable current lane-count observations.

In practice, lane sensing can fail locally even when most of an intersection remains observable. One or several lane counts may be unavailable while neighboring detectors continue to report valid measurements. If a missing count is represented by zero or a simple default, the controller acts on a distorted view of demand. Lane recovery is therefore a control problem as well as a data reconstruction problem {[}1{]}.

Signed Max-Pressure (SMP) is a well-established decentralized traffic signal controller {[}2,3{]}. It compares upstream and downstream lane states for every legal phase. An upstream count raises the relevant pressure contribution, while a downstream count lowers it.

Two properties of SMP motivate a probabilistic, movement-aware design. First, pressure is computed by summing over upstream and downstream lanes, and a single lane can contribute to several movements, so a reconstruction error can enter a phase score through multiple terms and change the selected phase when competing pressures are close. A point estimate---a single predicted value for a missing lane count---does not by itself quantify prediction uncertainty. A predictive standard deviation provides an explicit scale for adjusting reconstructed counts before pressure computation. Second, the movement relations that define pressure are also a natural structure for reconstruction: a lane-movement graph propagates information along these dependencies.

Research on incomplete sensing covers signal policies for partial observations, state restoration before control, and reconstruction of incomplete traffic data. MissLight {[}4{]} considers state imputation and a state-and-reward formulation for control, while RobustLight {[}7{]} restores inputs before an existing control platform. Temporal and graph-based imputation methods exploit available observations and dependencies among variables {[}10{]}--{[}12{]}, {[}17{]}--{[}21{]}. Pressure-based approaches also operate with estimated queues or partial connected-vehicle information {[}8{]}, {[}9{]}. Building on these foundations, this study examines how lane-level probabilistic reconstruction affects pressure scores, phase selection, and traffic performance when valid observations are preserved and the downstream controller is held fixed.

We propose a lane-movement probabilistic graph neural network (LMP-GNN) that estimates lane-count means and marginal standard deviations from lane-movement relations. Here, marginal refers to the predictive uncertainty for each lane considered separately; cross-lane covariance is not modeled. Three alternative rules convert these outputs into replacement counts. Predictive Mean uses the estimated mean directly. Fixed Lane Discount subtracts one predicted standard deviation and clips the result at zero. Staleness Gate reduces this uncertainty discount as the time since the last valid observation increases. Each rule replaces only unavailable counts; the resulting values are merged with the unchanged observed counts and supplied to SMP.

Selective reconstruction keeps every available measurement as a measured input. All compared reconstruction methods feed the same SMP controller. Using one common controller allows differences in pressure scores, phase choice, and traffic to be traced to the replacement values used for unavailable lanes. The completed lane-count vector can also be tested with other controllers that accept the same lane-level input.

Evaluation links three levels: reconstruction accuracy and marginal calibration, pressure-score accuracy and phase agreement at decision time, and closed-loop traffic performance. The selected lane-movement variant, labeled A2, has three residual lane-movement message-passing layers and no learned phase messages. Across 4,333,392 masked lane events, A2 achieved a mean absolute error (MAE) of 0.7873 vehicles per lane. Marginal coverage within one and two predicted standard deviations was 73.20\% and 94.56\%, respectively. Figure 9 reports complete-versus-reconstructed pressure-score and phase comparisons along each method's own closed-loop trajectory, using road-level mean imputation (Road Mean) as a baseline; the stored-trace audit is reported separately.

Across five fixed-demand CityFlow networks, Fixed Lane Discount reduced accrued average travel time (ATT) by up to 13.74\% relative to Road Mean under correlated missingness. It remained favorable at 60\% random missingness, whereas Road Mean performed better at 80\% and 90\%. On Gudang, Staleness Gate achieved lower accrued ATT than the decision-time adaptations of Bidirectional Recurrent Imputation for Time Series (BRITS) and the mask-aware graph imputation network (MagiNet) at 20\% and 40\% random and correlated missingness. MagiNet achieved lower accrued ATT at 60\% and 80\%, while approach-blackout comparisons remained unresolved.

This paper makes three contributions:

1. Lane-movement probabilistic reconstruction. LMP-GNN estimates lane-count means and marginal standard deviations from the movement relations used in pressure computation. A selective merge preserves every observed count and substitutes reconstructed values only for unavailable lanes.

2. Interpretable uncertainty-adjusted inputs. Predictive Mean, Fixed Lane Discount, and Staleness Gate convert the same model outputs into alternative replacement counts. Keeping SMP unchanged allows the effects of these rules to be compared through pressure scores, phase choices, and closed-loop traffic performance.

3. Quantify traffic benefits and computational cost. Evaluation connects reconstruction and decision-time diagnostics with closed-loop experiments on five CityFlow networks. Fixed Lane Discount reduces accrued ATT by up to 13.74\% relative to Road Mean under correlated missingness. The selected A2 model contains 63,362 parameters and has a median single-thread central processing unit (CPU) inference time of 0.983 ms. The decision-time BRITS and MagiNet adaptations use 9.43 and 29.81 times as many parameters and take 5.28 and 19.98 times as long for median inference under the same single-thread CPU benchmark, respectively.

\section{2. Related work}\label{related-work}

\subsection{2.1. Traffic signal control under missing observations}\label{traffic-signal-control-under-missing-observations}

One line of research asks whether traffic signal policies remain effective when observations degrade. Rodrigues and Lima Azevedo {[}13{]} evaluated deep reinforcement-learning control under demand surges, incidents, and sensor failures. RGLight {[}14{]} studied robustness and generalization when traffic flows, road networks, or observations change. These studies establish sensing degradation as an explicit operating condition for learned signal control.

Another line designs the policy around partial visibility. Zhang et al. {[}15{]} considered a vehicle-to-infrastructure setting in which only equipped vehicles are detected. BlindLight {[}16{]} addressed intersections that receive no local traffic information because detectors are absent or have failed. In both cases, the policy is developed to act on a partial view rather than receiving a separate reconstruction of selected lane measurements.

A third line of research combines state recovery with learned action selection. DiffLight {[}5{]} completes missing traffic information and generates signal decisions within one conditional diffusion framework. Xu et al. {[}6{]} first repair the missing agent state and then pass the completed state to a graph-based reinforcement-learning policy. These systems integrate restoration and action selection within their control frameworks.

In these control frameworks, traffic performance reflects both state recovery and the downstream decision policy. LMP-GNN addresses a different question: how replacement lane values affect pressure scores, phase choices, and traffic performance when every reconstruction method feeds the same SMP controller.

\subsection{2.2. State restoration for traffic-signal control}\label{state-restoration-for-traffic-signal-control}

State-restoration studies show that unavailable traffic state can be estimated before control. MissLight {[}4{]} reconstructs the current state of intersections without local observations from neighboring information. The completed state can support Max-Pressure, while a separate formulation also estimates state and reward for reinforcement learning.

RobustLight {[}7{]} places a diffusion-based restoration component before an existing traffic-signal-control platform. It repairs noisy, missing, or attacked platform state and passes the restored input to the downstream controller. This plug-in design preserves the existing control platform while adding state restoration.

These systems establish a broad restoration-to-control pipeline. The present setting is more selective: one or several lane detectors may fail while neighboring lanes at the same intersection continue to report valid counts. LMP-GNN retains those measurements, reconstructs the unavailable lane counts, uses the completed vector to calculate legal-phase pressures, and evaluates the traffic created by repeated phase decisions.

\subsection{2.3. General and traffic-specific imputation for reconstruction}\label{general-and-traffic-specific-imputation-for-reconstruction}

General imputation methods show how missing values can be recovered from temporal history and relationships among variables. BRITS uses recurrent models in both temporal directions {[}17{]}. GRIN adds graph relations, allowing one time series to draw information from connected series {[}18{]}. These methods demonstrate the value of history and cross-variable structure.

Probabilistic imputation describes a distribution of plausible missing values. CSDI, a conditional score-based diffusion model, conditions reconstruction on observed values {[}10{]}, while PriSTI strengthens conditional diffusion with spatiotemporal context {[}19{]}. These studies illustrate how conditional generative models can represent uncertainty in missing values. In a complementary direction, SPIN reconstructs sparse spatiotemporal observations through attention-based propagation over available measurements and graph relations {[}20{]}.

Traffic-specific methods add road-network structure and characteristic missingness patterns. ST-GIN combines graph-based spatial modeling, bidirectional temporal modeling, and uncertainty quantification {[}11{]}. MagiNet represents the missingness pattern while learning spatial and temporal relationships {[}12{]}. STAMImputer uses specialized processing paths and dynamic graph relations for block-wise missing traffic data {[}21{]}.

The information available to an offline imputer and a live signal controller is different. An offline model may use measurements collected after the missing value occurs. At decision \(t\), a signal controller has access only to observations from \(t\) and earlier decisions. The BRITS comparison therefore uses a forward-only adaptation, while the MagiNet comparison uses a trailing 12-decision window ending at \(t\). Neither adaptation receives the hidden count at \(t\) or any later observation.

These methods provide the temporal, relational, and probabilistic tools needed for reconstruction. For traffic control, the reconstructed value must also be evaluated after it enters the pressure calculation, affects phase choice, and changes later traffic. Section 2.5 brings these requirements together.

\subsection{2.4. Max-Pressure with estimated or partial state}\label{max-pressure-with-estimated-or-partial-state}

Max-Pressure control selects phases using upstream and downstream traffic states {[}2{]}, {[}3{]}. Under partial sensing, the source and representation of those states become part of the control design.

Backpressure control with estimated queue lengths (BP-EQ) {[}8{]} estimates a speed field from connected-vehicle observations, converts estimated speeds to densities and queue sizes, and supplies those estimates to backpressure control. Connected-vehicle Max-Pressure (CV-MP) {[}9{]} instead constructs pressure using connected-vehicle travel times in partially connected environments.

The present study examines missing lane-count measurements within otherwise partly observed intersections. Valid counts remain available to SMP, while probabilistic reconstruction supplies only missing entries. This setting allows the effects of replacement counts to be assessed through signed pressure scores, phase agreement, and closed-loop traffic outcomes.

\subsection{2.5. Research gaps and study positioning}\label{research-gaps-and-study-positioning}

The literature establishes four foundations. Traffic signal policies can operate with incomplete observations. Missing traffic state can be restored before control. Temporal, graph-based, and probabilistic methods can reconstruct incomplete data. Pressure-based control can also use estimated state.

Within the representative studies reviewed here, the effects of lane-wise predictive uncertainty on pressure scores, phase selection, and closed-loop traffic under an unchanged Signed Max-Pressure controller remain insufficiently characterized. This study addresses this gap by preserving valid observations and converting predictive means and standard deviations into replacement counts only for missing lanes.

LMP-GNN uses only information available at the current decision and combines lane-movement message passing with predictive means, marginal standard deviations, and three replacement rules. Evaluation follows the resulting counts from reconstruction accuracy and marginal calibration to pressure-score accuracy, phase agreement, and closed-loop traffic performance. Table I compares these design choices with those of representative studies.

\begin{table*}[!t]
\centering
\caption*{\textbf{Table I.} Positioning of representative studies and LMP-GNN}
\label{tab:I}

\begin{tblr}{width=\linewidth,colspec={Q[0.1400,l,m] Q[0.1600,l,m] Q[0.1500,l,m] Q[0.1050,c,m] Q[0.1600,l,m] Q[0.1800,l,m] Q[0.1050,c,m]},
colsep=3.4pt,rowsep=3pt,cells={font=\TableBodyFont,valign=m},
rows={ht=42pt},row{1}={font=\TableBodyFont\bfseries,halign=c,ht=25pt},
hline{1,Z}={0.6pt},hline{2}={0.35pt},
row{2}={font=\TableBodyFont\bfseries\itshape,ht=19pt,rowsep=3pt},
hline{2,3}={0.25pt},
row{5}={font=\TableBodyFont\bfseries\itshape,ht=19pt,rowsep=3pt},
hline{5,6}={0.25pt},
row{8}={font=\TableBodyFont\bfseries\itshape,ht=19pt,rowsep=3pt},
hline{8,9}={0.25pt},
row{11}={font=\TableBodyFont\bfseries\itshape,ht=19pt,rowsep=3pt},
hline{11,12}={0.25pt}}
Study & Missing state & Information timing & Uncertainty & Observed values & Controller & Closed-loop \\
\SetCell[c=7]{l} \emph{\textbf{Traffic signal control under missing observations}} &  &  &  &  &  &  \\
Zhang et al. {[}15{]} & Undetected vehicles & Current partial detections & No & Not a reconstruction interface & Learned RL & Yes \\
DiffLight {[}5{]} & Traffic state and reward & Offline learned framework & No lane-level output & No separate lane merge & Joint learned policy & Yes \\
\SetCell[c=7]{l} \emph{\textbf{State restoration before control}} &  &  &  &  &  &  \\
MissLight {[}4{]} & Intersection state and reward & Current neighboring state & No lane-level output & Restored intersection state & Max-Pressure or RL & Yes \\
RobustLight {[}7{]} & Platform traffic state & Online restored state & No lane-level output & Restored platform state & Existing platform & Yes \\
\SetCell[c=7]{l} \emph{\textbf{Imputation methods}} &  &  &  &  &  &  \\
BRITS {[}17{]} & Time-series entries & Bidirectional sequence & No & Not a controller interface & None & No \\
MagiNet {[}12{]} & Traffic measurements & Full-window published model & No & Not a controller interface & None & No \\
\SetCell[c=7]{l} \emph{\textbf{Pressure control with estimated or partial state}} &  &  &  &  &  &  \\
BP-EQ {[}8{]} & Estimated queue lengths & Current connected-vehicle state & No & Not a lane-count merge & Backpressure & Yes \\
CV-MP {[}9{]} & Connected-vehicle travel times & Current partial state & No & Not a lane-count merge & Travel-time Max-Pressure & Yes \\
\textbf{LMP-GNN (this study)} & Missing lane counts & Through the current decision & Yes, per lane & Available counts retained & Unchanged Signed Max-Pressure & Yes \\
\end{tblr}

\vspace{4pt}
\begin{minipage}{\textwidth}\fontsize{8.5}{10}\selectfont Note. Information timing summarizes whether a method uses current and past observations or a bidirectional or full-window procedure. Numerical results are not compared because the studies use different data, missingness processes, controllers, and metrics. RL denotes reinforcement learning, BP-EQ denotes backpressure control with estimated queue lengths, and CV-MP denotes connected-vehicle Max-Pressure. Bold identifies the present study.\end{minipage}
\end{table*}

\section{3. Traffic control setting and problem definition}\label{traffic-control-setting-and-problem-definition}

This section defines the traffic objects and control contract used throughout the study. It then defines missing lane observations, the information available at decision \(t\), and the componentwise merge that supplies a completed lane state to SMP.

\subsection{3.1. Traffic network, movements, and signal phases}\label{traffic-network-movements-and-signal-phases}

Consider a traffic network with signalized intersections \(j\) \ensuremath{\in} \(\mathcal{I}\) and lanes \(i\) \ensuremath{\in} \(\mathcal{L}\), where \(\mathcal{I}\) and \(\mathcal{L}\) denote the sets of signalized intersections and lanes, respectively. A movement \(r\) \ensuremath{\in} \(\mathcal{R}\) connects one road approaching the intersection to one road leaving it. In the CityFlow road network file, this road-level movement is represented as a roadLink.

A roadLink contains one or more laneLinks. Each laneLink (\(u\), \(d\)) is a concrete connection from incoming lane \(u\) to outgoing lane \(d\). RoadLinks describe permitted road-to-road movements, while laneLinks describe the lane-level paths that realize them. Figure 1 illustrates this hierarchy.

\begin{figure*}[!t]

\centering

\includegraphics[width=1.0\textwidth]{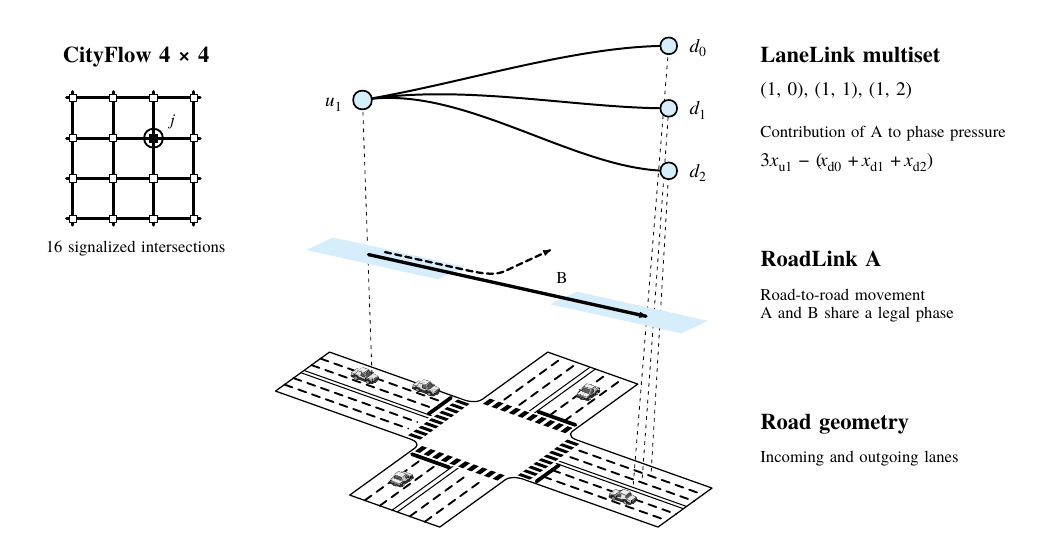}

\caption{Traffic objects at a CityFlow 4 × 4 intersection: incoming/outgoing lanes, roadLinks, and concrete laneLinks. In the illustrated legal phase, roadLinks A and B are released together; the repeated contribution follows A's actual laneLinks.}

\label{fig:1}

\end{figure*}

At intersection \(j\), \(\Phi_{j}\) is the set of legal phases and \(p\) \ensuremath{\in} \(\Phi_{j}\) denotes one phase. For each nonempty phase, \(\mathcal{E}_{j,p}\) is the multiset of released laneLinks. Repeated lane appearances are retained because every connection contributes separately to the pressure score. The index \(t\) denotes a signal decision and \(q_{t,i}\) denotes the true count on lane \(i\).

\begin{table*}[!t]
\centering
\caption*{\textbf{Table II.} Core notation for the traffic-control and reconstruction formulation}
\label{tab:II}
\begin{minipage}[t]{0.484\textwidth}
\vspace{0pt}

\begin{tblr}{width=\linewidth,colspec={Q[0.2050,l,m] Q[0.7950,l,m]},
colsep=3.4pt,rowsep=3pt,cells={font=\TableBodyFont,valign=m},
rows={ht=21pt},row{1}={font=\TableBodyFont\bfseries,halign=c,ht=25pt},
hline{1,Z}={0.6pt},hline{2}={0.35pt},
row{2}={font=\TableBodyFont\bfseries\itshape,ht=19pt,rowsep=3pt},
hline{2,3}={0.25pt},
row{7}={font=\TableBodyFont\bfseries\itshape,ht=19pt,rowsep=3pt},
hline{7,8}={0.25pt},
row{11}={font=\TableBodyFont\bfseries\itshape,ht=19pt,rowsep=3pt},
hline{11,12}={0.25pt}}
Symbol & Definition \\
\SetCell[c=2]{l} \emph{\textbf{Traffic network}} &  \\
\(\mathcal{I}\), \(j\) & Set of signalized intersections; \(j\) indexes one intersection. \\
\(\mathcal{L}\), \(i\) & Set of lanes; \(i\) indexes one lane. \\
\(\mathcal{R}\), \(r\) & Set of road-level movements; \(r\) indexes one movement, represented as a CityFlow roadLink. \\
(\(u\), \(d\)) & Concrete laneLink from incoming lane \(u\) to outgoing lane \(d\). \\
\SetCell[c=2]{l} \emph{\textbf{Movements and legal signal phases}} &  \\
\(\Phi_{j}\), \(p\) & Legal phase set at intersection \(j\); \(p\) denotes one phase. \\
\(\mathcal{E}_{j,p}\) & Multiset of concrete laneLinks released by phase \(p\) at intersection \(j\). \\
\(t\) & Signal-decision index. \\
\SetCell[c=2]{l} \emph{\textbf{Traffic state and Signed Max-Pressure}} &  \\
\(q_{t,i}\) & True vehicle count on lane \(i\) at decision \(t\). \\
\(x_{t}\) & Completed lane-count vector supplied to Signed Max-Pressure. \\
\(P_{j,p}\)(\(x_{t}\)) & Signed pressure score of legal phase \(p\) at intersection \(j\). \\
\(a_{t,j}\) & Phase selected at intersection \(j\) at decision \(t\). \\
\end{tblr}

\end{minipage}
\hfill
\begin{minipage}[t]{0.484\textwidth}
\vspace{0pt}

\begin{tblr}{width=\linewidth,colspec={Q[0.2050,l,m] Q[0.7950,l,m]},
colsep=3.4pt,rowsep=3pt,cells={font=\TableBodyFont,valign=m},
rows={ht=21pt},row{1}={font=\TableBodyFont\bfseries,halign=c,ht=25pt},
hline{1,Z}={0.6pt},hline{2}={0.35pt},
row{2}={font=\TableBodyFont\bfseries\itshape,ht=19pt,rowsep=3pt},
hline{2,3}={0.25pt},
row{8}={font=\TableBodyFont\bfseries\itshape,ht=19pt,rowsep=3pt},
hline{8,9}={0.25pt}}
Symbol & Definition \\
\SetCell[c=2]{l} \emph{\textbf{Missing observations and stored lane history}} &  \\
\(Q_{t,i}\) & Stored lane-count value: observed count when available and zero when unavailable. \\
\(z_{t,i}\) & Availability label: \(z\) = 0 means observed and \(z\) = 1 means unavailable. \\
\(h_{t,i}\) & Most recently observed count stored for lane \(i\). \\
\(s_{t,i}\) & Number of signal decisions since lane \(i\) was last observed. \\
\(\mathcal{F}_{t}\) & Information available when the decision at time \(t\) is formed. \\
\SetCell[c=2]{l} \emph{\textbf{Reconstruction outputs and completed lane state}} &  \\
\(\mu_{t,i}\) & Predictive mean lane count for lane \(i\) at decision \(t\). \\
\(\sigma_{t,i}\) & Marginal predictive standard deviation for lane \(i\) at decision \(t\). \\
\(g_{k}\) & Missing-lane input rule \(k\), defined in Section 4.5. \\
\(x_{t,i}^{(k)}\) & Completed value for lane \(i\) supplied to the controller under rule \(k\). \\
\end{tblr}

\end{minipage}
\end{table*}

\subsection{3.2. Signed Max-Pressure control}\label{signed-max-pressure-control}

The downstream decision rule is the ordinary SMP controller used throughout the study {[}2,3{]}. Let \(x_{t}\) denote the completed lane count vector supplied to the controller. Under complete observation, \(x_{t}\) contains measured counts. Under missing observations, Section 3.4 defines how measured and reconstructed values are merged.

For every legal nonempty phase \(p\), SMP sums the upstream count minus the downstream count for each released laneLink:

\setcounter{equation}{0}
\begin{equation}
P_{j,p}\left( x_{t} \right)=\sum_{(u,d) \in E_{j,p}}^{}\left( x_{t,u}-x_{t,d} \right).
\label{eq:1}
\end{equation}

Each laneLink (\(u\), \(d\)) contributes \(x_{t,u}\) \ensuremath{-} \(x_{t,d}\). A larger upstream count raises the phase score, while a larger downstream count lowers it. The sum uses the full multiset \(\mathcal{E}_{j,p}\), so a lane that appears in several laneLinks contributes once per connection. One lane error can therefore change several pressure terms and reorder competing phases.

SMP then selects the legal phase with the largest pressure score:

\setcounter{equation}{1}
\begin{equation}
a_{t,j}=\arg\max_{p \in \Phi_{j}:E_{j,p} \neq \varnothing}P_{j,p}\left( x_{t} \right),
\label{eq:2}
\end{equation}

When several legal phases share the largest score, the controller selects the first maximum listed in the CityFlow road network. The same completed state therefore produces the same phase choice.

\subsection{3.3. Missing lane observations and decision-time information}\label{missing-lane-observations-and-decision-time-information}

A missing lane observation means that the current lane count is unavailable to both the reconstructor and the controller, while the physical lane remains open to traffic. At decision \(t\), reconstruction uses only information that has already been received. The hidden count of the unavailable lane and all later measurements remain outside the input.

Let \(q_{t}\) be the true lane-count vector at decision \(t\), and let \(z_{t}\) \ensuremath{\in} \({\{ 0,\ 1\}}^{|\mathcal{L}|}\) be the binary missingness-mask vector, where \textbar{}\(\mathcal{L}\)\textbar{} is the number of lanes. An entry \(z_{t,i}\) = 1 indicates an unavailable count, whereas \(z_{t,i}\) = 0 indicates an observed count. In the experiments, the prescribed missing-observation process generates \(z_{t}\) and supplies it as an availability indicator. Missingness is therefore specified by the mask, not inferred from the numerical value of \(q_{t}\),\textsubscript{i}.

\setcounter{equation}{2}
\begin{equation}
Q_{t,i}=\left( 1-z_{t,i} \right)q_{t,i}.
\label{eq:3}
\end{equation}

The stored count vector is \(Q_{t}\) = (1 \ensuremath{-} \(z_{t}\)) \ensuremath{\odot} \(q_{t}\), where \ensuremath{\odot} denotes elementwise multiplication, as defined in (3). Thus, \(Q_{t,i}\) equals \(q_{t,i}\) for an observed lane and zero for an unavailable lane. Providing \(Q_{t}\) and \(z_{t}\) separately distinguishes an observed zero count from a missing value stored as zero. For masked lanes, the simulator count \(q_{t,i}\) is retained as an evaluation target and is not supplied to the reconstructor or controller.

Let \(h_{t,i}\) be the most recently observed count for lane \(i\) and let \(s_{t,i}\) be the number of signal decisions since that observation. These values are carried into the reconstruction formed at decision \(t\). After the current observation is processed, an observed lane updates \(h\) and resets \(s\) for the next decision. An unavailable lane retains \(h\) and its staleness increases for the next decision.

The decision-time information set \(\mathcal{F}_{t}\) contains the fixed road network, all observations and availability labels received no later than \(t\), the mask history, the last observed vector \(h_{t}\), and the staleness vector \(s_{t}\). It excludes the hidden count of a masked lane and every future measurement. Figure 2 illustrates the temporal part of this boundary.

\begin{figure*}[!t]

\centering

\includegraphics[width=1.0\textwidth]{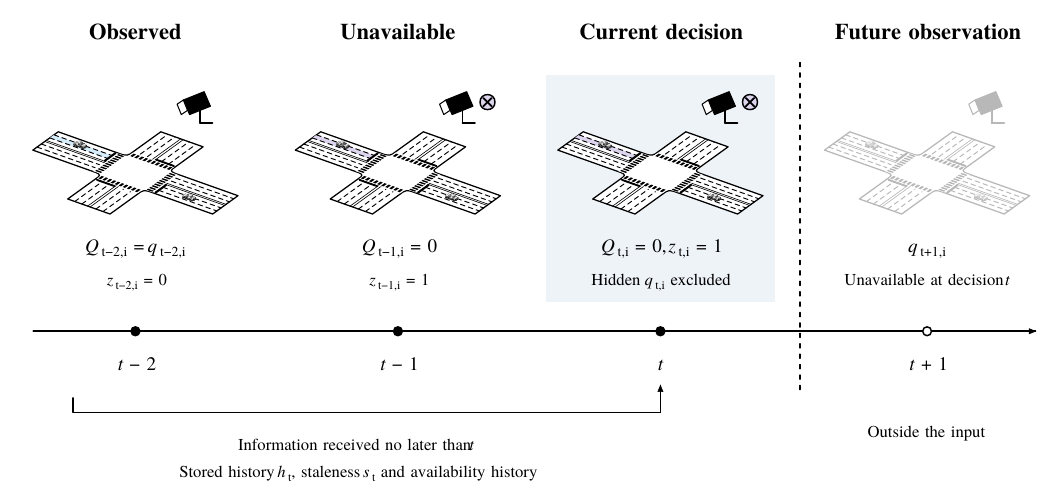}

\caption{Information available at decision t. Reconstruction uses received observations and missingness history up to t; hidden counts and future observations remain unavailable.}

\label{fig:2}

\end{figure*}

\subsection{3.4. From reconstructed lane counts to signal decisions}\label{from-reconstructed-lane-counts-to-signal-decisions}

For each lane, LMP-GNN produces a predictive mean \(\mu_{t,i}\) and a marginal predictive standard deviation \(\sigma_{t,i}\). The corresponding marginal variance is \(v_{t,i}\) = \(\sigma_{t,i}^{2}\). Only outputs for unavailable lanes are used to replace counts; observed counts pass through unchanged. Section 4 describes the model outputs, and Section 4.5 defines the three replacement rules.

The controller requires one completed value for every lane used in its phase scores. Available measurements pass directly into the completed state. Reconstruction is used only when the current measurement is unavailable.

Let \(g_{k}\) denote missing-lane input rule \(k\). The value supplied for lane \(i\) is:

\setcounter{equation}{3}
\begin{equation}
x_{t,i}^{(k)}=(1 - z_{t,i})q_{t,i}+z_{t,i}g_{k}(\mu_{t,i},\sigma_{t,i},s_{t,i})
\label{eq:4}
\end{equation}

When \(z_{t,i}\) = 0, the observed count \(q_{t,i}\) passes through unchanged. When \(z_{t,i}\) = 1, rule \(g_{k}\) supplies the missing lane value. The resulting vector \(x_{t}^{(k)}\) is then evaluated by the SMP score and phase selection rules in Section 3.2. Section 4.5 defines the three input rules used in the study.

Figure 3 summarizes the complete handoff from sensing to reconstruction and control. Section 4 then describes the internal LMP-GNN architecture.

\begin{figure*}[!t]

\centering

\includegraphics[width=1.0\textwidth]{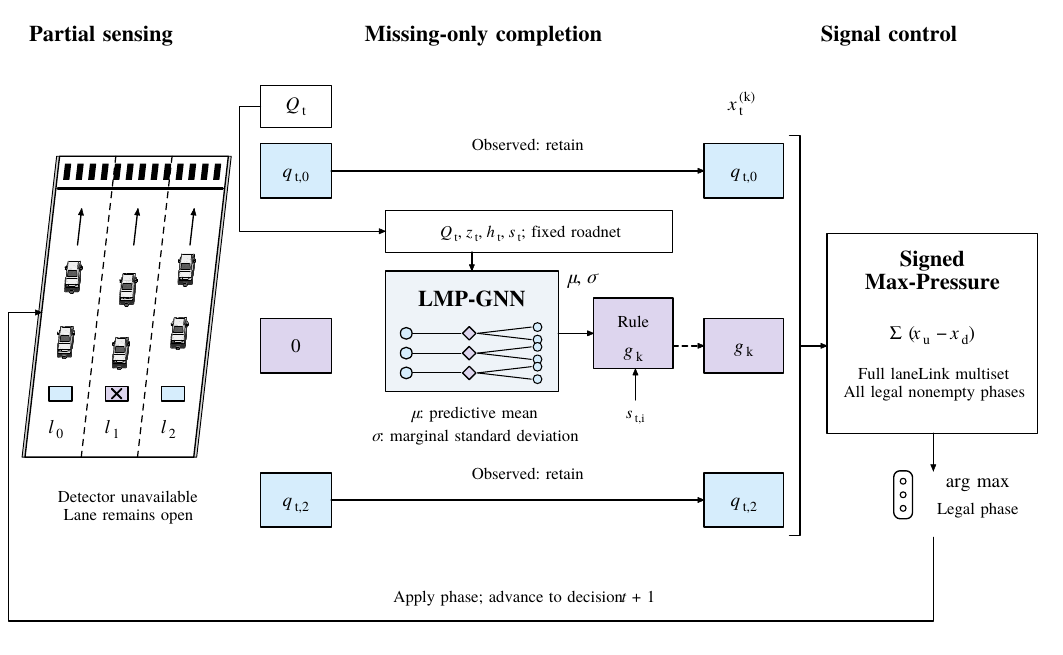}

\caption{Reconstruction and control handoff. Observed lane counts pass through unchanged; only missing counts are replaced using LMP-GNN's predictive mean and marginal standard deviation. The completed counts enter ordinary Signed Max-Pressure, which selects a legal phase.}

\label{fig:3}

\end{figure*}

Two complementary evaluations follow this handoff. First, decision-time diagnostics compare reconstructed-input pressure scores and phase choices with complete-count references at the same traffic state. Figure 9 computes these diagnostics along each method\textquotesingle s own closed-loop trajectory. Second, a closed-loop evaluation applies the selected phases and advances the simulator, showing how repeated decision differences affect later queues and accumulated travel time. Both are needed because the same count error can affect SMP differently depending on whether the lane is upstream or downstream, how many laneLinks contain it, and how close the leading phase scores are.

\section{4. LMP-GNN reconstruction method}\label{lmp-gnn-reconstruction-method}

\subsection{4.1. LMP-GNN architecture}\label{lmp-gnn-architecture}

LMP-GNN places a reconstruction stage between decision-time sensing and the SMP controller. At decision \(t\), the model receives stored lane values, availability labels, last-observed counts, staleness, and lane-movement relations from the road network. Every input satisfies the information boundary defined in Section 3.3.

Each lane is represented by 12 features and encoded into a hidden state (Figure 4). Three lane-movement message-passing layers combine information from lanes connected through the same roadLink movements. A shared output head produces a predictive mean and marginal standard deviation for every lane. For each unavailable lane, Section 4.5 applies Predictive Mean, Fixed Lane Discount, or Staleness Gate to produce a replacement count. The observed and reconstructed counts form the completed lane-count vector passed to SMP. The road network and SMP continue to define legal phases, complete laneLink multiplicity, pressure calculation, tie handling, and phase selection.

\begin{figure*}[!t]

\centering

\includegraphics[width=1.0\textwidth]{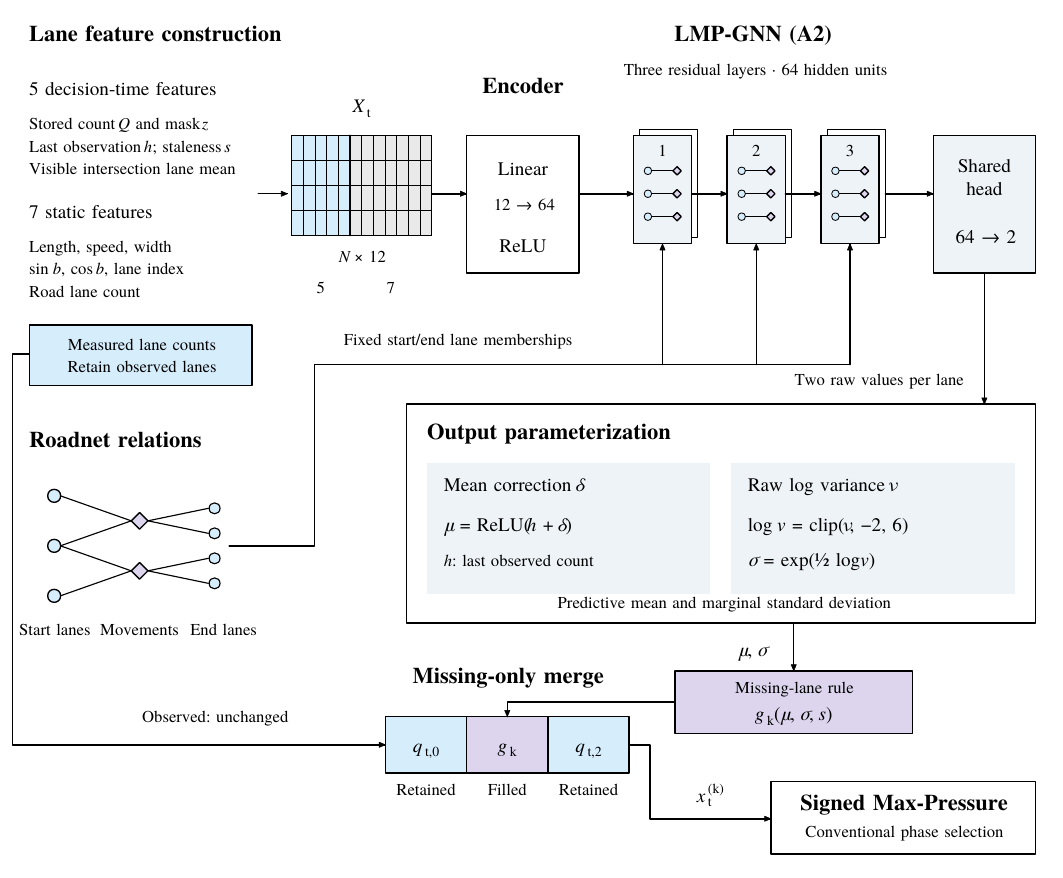}

\caption{A2 architecture and missing-only merge. Twelve lane features feed three lane-movement layers and a shared mean/variance head. Observed counts remain unchanged; reconstructed missing counts complete the input to Signed Max-Pressure.}

\label{fig:4}

\end{figure*}

\subsection{4.2. Lane and movement graph and input features}\label{lane-and-movement-graph-and-input-features}

A lane history becomes less informative as the last observation grows stale. Lanes connected through the same movement provide complementary information about demand entering the movement and the downstream state that receives it. The road network provides these relations directly, so reconstruction uses traffic context that is specific to the missing lane rather than a generic spatial neighborhood.

The roadnet extractor represents each road lane as a lane node and each roadLink as a movement relation between its start and end lanes. A2 uses these lane-movement relations to reconstruct lane counts, but it does not learn or change which movements belong to a legal phase. The road-network file defines phase membership, and SMP later scores each phase using every concrete laneLink it releases. If one incoming lane connects to three outgoing lanes released by the same phase, its completed count enters three pressure terms, so one reconstruction error can affect that phase score three times. A3 adds learned phase messages only as a capacity-matched architecture comparison.

\setcounter{equation}{4}
\begin{equation}
x_{t,i}=\begin{pmatrix}
\log(1+Q_{t,i}),z_{t,i},log(1+h_{t,i}),\\min(s_{t,i},50)/50,log(1+{\acute{q}}_{t,\iota(i)}^{vis}),\\\mathcal{l}_{i}/1000,v_i^{\max}/20,w_i/5,\\sinb_i,cosb_i,k_i/5,n_i^{road}/5
\end{pmatrix}.
\label{eq:5}
\end{equation}

The first five features are the stored count, availability label, last observed count, normalized staleness, and visible lane mean at the associated intersection. The remaining seven features describe lane length, speed limit, width, travel bearing, lane index, and road lane count. A missing observation is stored as zero while its availability label remains one, so the hidden target never enters feature construction.

\subsection{4.3. Probabilistic reconstruction model}\label{probabilistic-reconstruction-model}

A linear encoder first maps each lane feature vector to a 64-dimensional hidden representation:

\setcounter{equation}{5}
\begin{equation}
h_{i}^{(0)}=ReLU\left( W_{enc}x_{t,i}+b_{enc} \right).
\label{eq:6}
\end{equation}

Each A2 layer then forms separate summaries of the start-lane and end-lane representations for every movement (Figure 5):

\setcounter{equation}{6}
\begin{equation}
\begin{gathered}{\acute{h}}_{r,S}^{\left( \mathcal{l} \right)}={mean}_{i \in S(r)}h_{i}^{\left( \mathcal{l} \right)},\\{\acute{h}}_{r,D}^{\left( \mathcal{l} \right)}={mean}_{i \in D(r)}h_{i}^{\left( \mathcal{l} \right)}.\end{gathered}
\label{eq:7}
\end{equation}

Separate affine transformations combine the two summaries into one movement representation:

\setcounter{equation}{7}
\begin{equation}
m_{r}^{\left( \mathcal{l} \right)}=ReLU\left( W_{S}^{\left( \mathcal{l} \right)}{\acute{h}}_{r,S}^{\left( \mathcal{l} \right)}+W_{D}^{\left( \mathcal{l} \right)}{\acute{h}}_{r,D}^{\left( \mathcal{l} \right)} \right).
\label{eq:8}
\end{equation}

Every learned affine transformation includes a bias, which is omitted from Eqs. (7) to (10) for readability. For each lane, the model gathers the representations of movements that start from that lane and movements that end at that lane. It averages the two groups separately to produce two summaries of the lane\textquotesingle s movement context:

\setcounter{equation}{8}
\begin{equation}
\begin{gathered}d_{i}^{\left( \mathcal{l} \right)}={mean}_{r:i \in S(r)}m_{r}^{\left( \mathcal{l} \right)},\\u_{i}^{\left( \mathcal{l} \right)}={mean}_{r:i \in D(r)}m_{r}^{\left( \mathcal{l} \right)}.\end{gathered}
\label{eq:9}
\end{equation}

The model combines the lane\textquotesingle s current representation with the two movement-context summaries. It then adds the resulting update to the existing lane representation instead of replacing it. This direct carryover is called a skip connection, and the sum forms the representation passed to the next layer:

\setcounter{equation}{9}
\begin{equation}
\begin{aligned}h_{i}^{\left( \mathcal{l}+1 \right)}=h_{i}^{\left( \mathcal{l} \right)}{}&+ReLU\bigl( W_{self}^{\left( \mathcal{l} \right)}h_{i}^{\left( \mathcal{l} \right)}\\&\quad+W_{start}^{\left( \mathcal{l} \right)}d_{i}^{\left( \mathcal{l} \right)}+W_{end}^{\left( \mathcal{l} \right)}u_{i}^{\left( \mathcal{l} \right)} \bigr).\end{aligned}
\label{eq:10}
\end{equation}

\begin{figure*}[!t]

\centering

\includegraphics[width=1.0\textwidth]{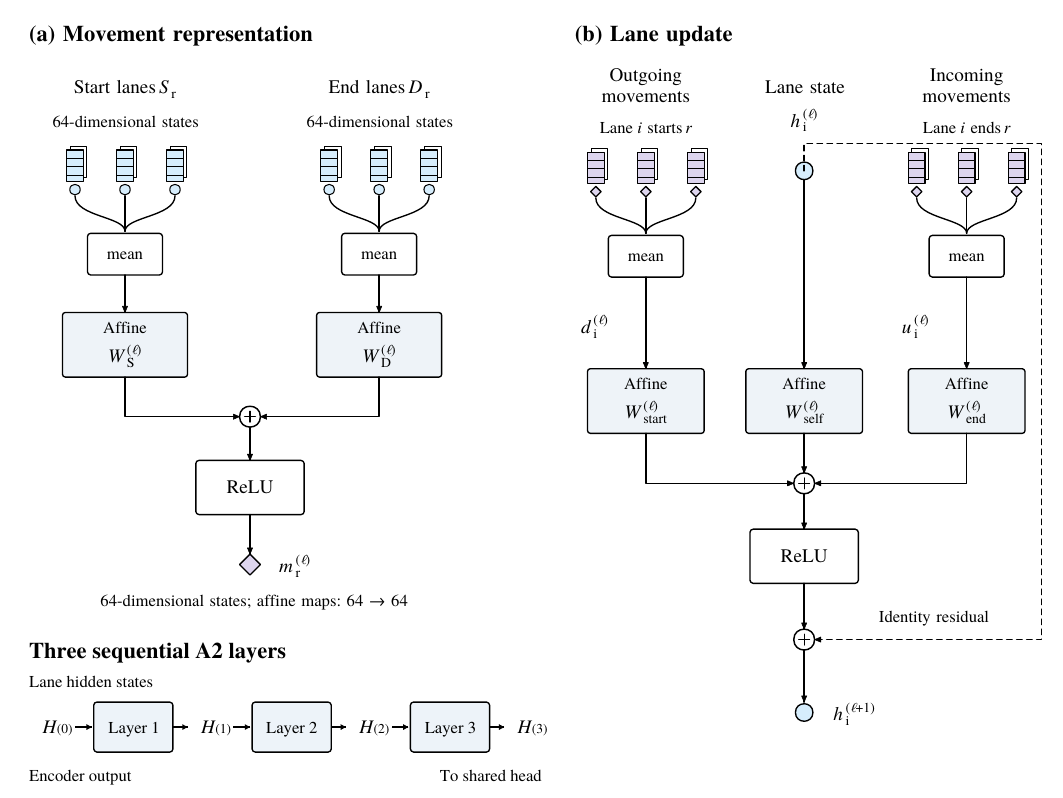}

\caption{A2 lane-movement message passing: (a) separate start/end-lane aggregation forms movement representations; (b) incoming/outgoing movement summaries update lane states through a residual connection. Three layers have separate parameters. \(H^{(\ell)}\) collects the individual lane states \(h_i^{(\ell)}\).}

\label{fig:5}

\end{figure*}

The update is repeated for three layers with layer-specific parameters. The hidden representation is different from the scalar last-observed count defined in Section 3.3. A shared two-output head maps the final representation to a mean correction and a raw log variance:

\setcounter{equation}{10}
\begin{equation}
\begin{gathered}\mu_{t,i}=ReLU\left( h_{t,i}+\delta_{t,i} \right),\\ logv_{t,i}=clip\left( {\overset{\sim}{v}}_{t,i}, - 2,6 \right),\\ \sigma_{t,i}=\exp\left( \frac{1}{2}\log v_{t,i} \right).\end{gathered}
\label{eq:11}
\end{equation}

The predictive mean is anchored to the stored last observation. The reported uncertainty is marginal for each lane and excludes cross-lane covariance. With input width 12, hidden width 64, and three A2 layers, the retained checkpoint contains 63,362 trainable parameters.

\subsection{4.4. Model training}\label{model-training}

Training uses clean CityFlow SMP traces generated without artificial missing observations. Artificial masks hide selected lane counts, and those hidden values remain available only as reconstruction targets. The model receives the masked stored counts and availability labels, and the loss is computed only at masked positions:

\setcounter{equation}{11}
\begin{equation}
L(\theta)=\frac{1}{|\Omega)}\sum_{(t,i) \in \Omega}^{}\frac{1}{2}\left\lbrack \log v_{t,i}+\frac{\left( q_{t,i} - \mu_{t,i} \right)^{2}}{v_{t,i}} \right).
\label{eq:12}
\end{equation}

The objective trains the predictive mean and marginal uncertainty jointly. Weight updates use five synthetic CityFlow scenarios. Gudang is used only for validation and early stopping, while the Heterogeneous network is excluded from both training and model selection. All primary results condition on one frozen A2 checkpoint.

\subsection{4.5. Uncertainty-adjusted reconstruction rules}\label{uncertainty-adjusted-reconstruction-rules}

The merge in Section 3.4 requires one controller-input value for every unavailable lane. The three rules in Eqs. (13) to (15) convert the predictive outputs into that value (Figure 6):

\setcounter{equation}{12}
\begin{equation}
g_{\mu}(\mu,\sigma,s)=\mu
\label{eq:13}
\end{equation}

\setcounter{equation}{13}
\begin{equation}
g_{fix}(\mu,\sigma,s)=\max(0,\mu - \lambda\sigma).
\label{eq:14}
\end{equation}

\setcounter{equation}{14}
\begin{equation}
g_{stale}(\mu,\sigma,s)=\max\left( 0,\mu - \lambda e^{- \frac{s}{\tau}}\sigma \right).
\label{eq:15}
\end{equation}

Predictive Mean uses the estimated mean directly. Fixed Lane Discount subtracts one predicted standard deviation and clips the value at zero. Staleness Gate applies the same discount after a recent observation loss and relaxes toward Predictive Mean as staleness increases. Observed values remain measured inputs. Because SMP subtracts downstream counts, discounting a reconstructed downstream value can raise a phase score. The three rules therefore transform the same probabilistic prediction in different ways before SMP calculates its legal-phase scores.

\begin{figure*}[!t]

\centering

\includegraphics[width=1.0\textwidth]{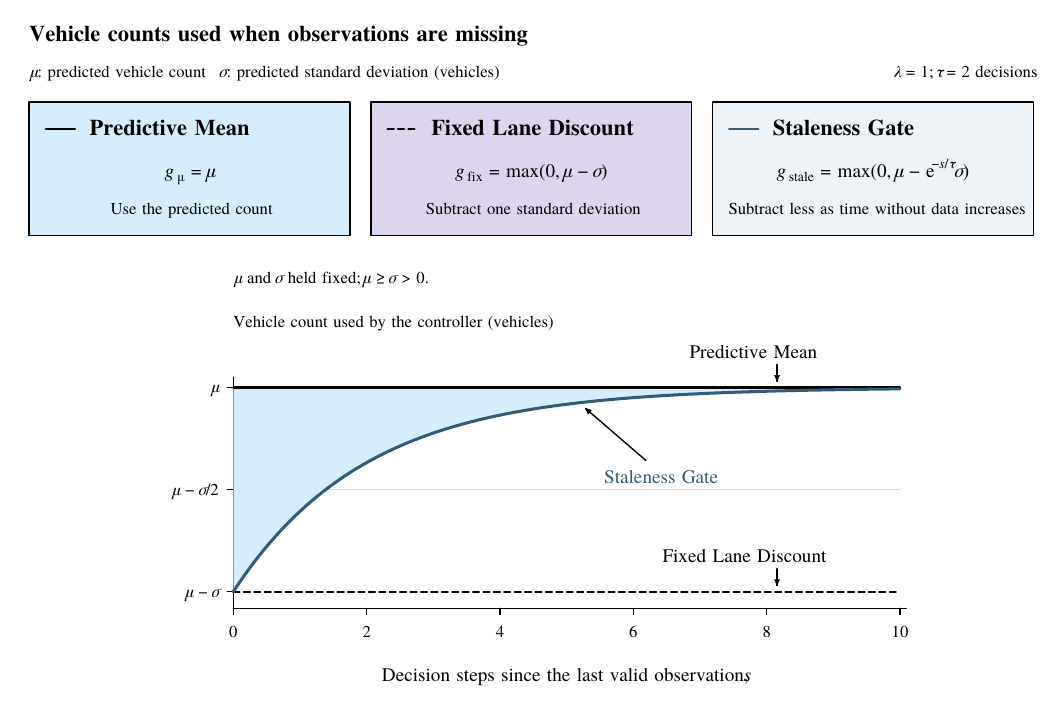}

\caption{Three missing-count rules, with \ensuremath{\lambda} = 1 and \ensuremath{\tau} = 2 decisions. For fixed \ensuremath{\mu} \ensuremath{\geq} \ensuremath{\sigma} \textgreater{} 0, Staleness Gate approaches Predictive Mean as time since observation increases. Blue shading shows the count discount, not a confidence interval.}

\label{fig:6}

\end{figure*}

\section{5. Experimental setup}\label{experimental-setup}

This section defines the simulation settings, missing-observation patterns, comparison methods, and evaluation metrics. It also explains what is repeated in each study and what therefore counts as one unit of analysis. In the fixed-demand study, one unit is a matched outage schedule applied to the same demand. In the demand-variation and geometry studies, one unit is a generated demand case after its matched mask runs are combined. Keeping these units separate prevents the analysis from counting every simulation row as an independent piece of statistical evidence.

\subsection{5.1. Simulation platforms and benchmark design}\label{simulation-platforms-and-benchmark-design}

CityFlow {[}22{]} is the primary platform because it repeatedly links lane observations, reconstruction, SMP phase choice, and traffic evolution. A fixed-state calculation stops at one stored traffic snapshot. It can show how a reconstructed input changes the current pressure scores and selected phase, but it cannot show what happens after that phase is applied. CityFlow advances traffic after every decision, so the selected phases change later queues, later observations, and accumulated travel time. Every CityFlow run uses a 1 s simulation step, a 3,600 s horizon, and a signal decision every 10 s. SUMO {[}23{]} is reserved for the auxiliary Cologne1 transfer study, and its native outcomes remain separate from the CityFlow measures.

The five CityFlow benchmarks differ in network structure, demand input, and experimental role (Table III). Synthetic 2 × 2, Synthetic 4 × 4, and Arterial 1 × 6 use synthetic demand. Gudang supplies validation and early stopping as well as evaluation, but it does not contribute training gradients. The Heterogeneous network is excluded from training, validation, early stopping, and model selection. Its mixture of four-arm and three-arm intersections tests whether the selected A2 model transfers to a structurally unseen network.

\begin{table*}[!t]
\centering
\caption*{\textbf{Table III.} CityFlow benchmark networks and experimental roles}
\label{tab:III}

\begin{tblr}{width=\linewidth,colspec={Q[0.1500,l,m] Q[0.1900,l,m] Q[0.1300,l,m] Q[0.0650,c,m] Q[0.0700,c,m] Q[0.0700,c,m] Q[0.3250,l,m]},
colsep=3.4pt,rowsep=3pt,cells={font=\TableBodyFont,valign=m},
rows={ht=42pt},row{1}={font=\TableBodyFont\bfseries,halign=c,ht=25pt},
hline{1,Z}={0.6pt},hline{2}={0.35pt}}
\textbf{Network} & \textbf{Structure} & \textbf{Demand} & \textbf{Signals} & \textbf{Eligible lanes} & \textbf{All lanes} & \textbf{Role} \\
Synthetic 2 × 2 & Regular 2 × 2 grid & Fixed synthetic & 4 & 48 & 72 & Training and primary evaluation \\
Synthetic 4 × 4 & Regular 4 × 4 grid & Fixed synthetic & 16 & 192 & 240 & Training, primary evaluation, and demand variation \\
Arterial 1 × 6 (700) & Six-intersection corridor & Fixed synthetic & 6 & 72 & 114 & Training, primary evaluation, and demand variation \\
Gudang & 4 × 4; 16 four-arm intersections & Gudang flow & 16 & 192 & 240 & Validation, early stopping, primary evaluation, and demand variation \\
\textbf{Heterogeneous} & 4 × 4; 12 four-arm, 4 three-arm intersections & Held-out demand & 16 & 168 & 228 & Structurally unseen evaluation; no training or selection \\
\end{tblr}

\end{table*}

The eligible-lane columns in Table III show how many lanes each missingness generator is allowed to hide. Correlated masks use movement-start lanes at eligible approaches, while random masks may use any roadnet lane. These counts describe masking eligibility, not vehicle demand.

The main missingness-severity experiment evaluates random and correlated missingness from 10\% to 90\% on five networks while holding each network\textquotesingle s demand file fixed. An outage schedule is the complete time sequence showing which lane observations are unavailable at every signal decision. A candidate and its comparator receive the same schedule, and their ATT difference is calculated within that matched schedule before the ten schedule-level differences are summarized.

A2 is trained on five synthetic scenarios, selected using Gudang validation, and evaluated with frozen weights. A separate demand-variation study changes the traffic realization on Synthetic 4 × 4, Arterial 1 × 6, and Gudang to determine whether two selected result directions persist beyond one fixed demand file.

All matched comparisons share the network, demand, horizon, legal phases, SMP equation, tie rule, and missing-observation schedule. Methods differ only in the value supplied for an unavailable lane. The retained A2 pipeline uses 12 features, hidden width 64, three message-passing layers, masked Gaussian negative log-likelihood, Adam with learning rate 0.001, batch size 64, gradient-norm clipping at 5, and at most 3,000 updates. Gudang validation is checked every 150 updates with patience 8. CityFlow uses one engine thread with lane changing and replay disabled. The software environment is Python 3.9.6, PyTorch 2.8.0, SciPy 1.13.1, and SUMO 1.27.1 through TraCI.

Each experiment repeats a different source of variation. The main experiment repeats matched outage schedules, the demand experiment repeats generated demand cases, the geometry experiment repeats demand-level geometry contrasts, the architecture comparison repeats model seeds, and the SUMO study repeats simulator schedules. These observations are analyzed separately.

\subsection{5.2. Missing observation scenarios}\label{missing-observation-scenarios}

The four scenarios differ in what becomes unavailable and how long the loss persists, as illustrated in Figure 7. Random missingness samples individual roadnet lanes at each decision. Correlated missingness selects incoming approach groups and hides their eligible movement-start lanes together. Sustained outage keeps one movement-start lane unavailable for 300 or 600 s. Approach blackout removes one eligible approach for the full run. In every case, the physical lanes remain open to traffic.

\begin{table*}[!t]
\centering
\caption*{\textbf{Table IV.} Implemented missing-observation scenarios and their masking contracts}
\label{tab:IV}

\begin{tblr}{width=\linewidth,colspec={Q[0.1600,l,m] Q[0.1550,l,m] Q[0.1700,l,m] Q[0.1650,l,m] Q[0.1700,l,m] Q[0.1800,l,m]},
colsep=3.4pt,rowsep=3pt,cells={font=\TableBodyFont,valign=m},
rows={ht=39pt},row{1}={font=\TableBodyFont\bfseries,halign=c,ht=25pt},
hline{1,Z}={0.6pt},hline{2}={0.35pt}}
Scenario & Selected unit & Eligible lane set & Spatial rule & Temporal rule & Executed condition \\
Random lane missingness & Individual lane & All roadnet lanes & Lanes sampled independently & Resampled at each decision & 10\%--90\% in 10-point steps \\
Correlated approach missingness & Incoming approach group & Eligible movement-start lanes & Selected group masked together & Independent group draw at each decision & 10\%--90\% in 10-point steps \\
Sustained lane outage & One movement-start lane & Eligible movement-start lanes & One lane selected & Unavailable continuously & 300 s and 600 s \\
Approach blackout & One incoming approach & Eligible lanes on that approach & All eligible lanes masked & Full 3,600 s run & Categorical condition \\
\end{tblr}

\end{table*}

For random missingness, 60\% means that each eligible lane has a 0.60 probability of being hidden at a decision. For correlated missingness, the same probability is applied to each eligible approach group, and selecting a group hides all of its eligible lanes together. The same nominal percentage can therefore produce a different number and spatial arrangement of unavailable lanes because it is applied to different masking units.

Each eligible lane in random missingness, or each eligible approach group in correlated missingness, receives one random number for a given network, schedule, and decision. A lane or group is hidden at 10\% when its number is below 0.10, at 20\% when it is below 0.20, and so on. A lane or approach group hidden at a lower severity therefore remains hidden at every higher severity in the same schedule. Every compared method receives exactly the same mask for the same network, demand, schedule, and severity.

\begin{figure*}[!t]

\centering

\includegraphics[width=1.0\textwidth]{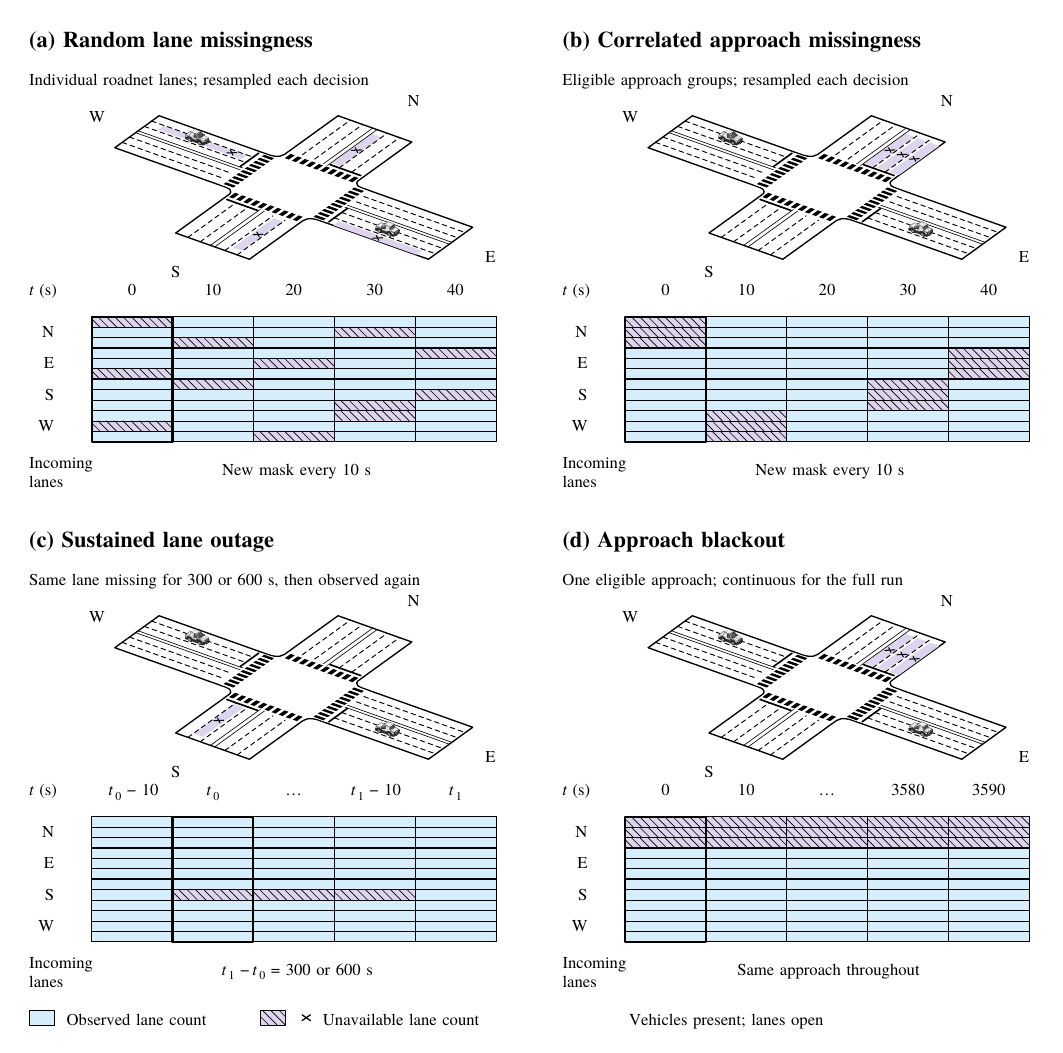}

\caption{Four observation-loss patterns: individual-lane random missingness, grouped movement-start-lane correlated missingness, one-lane sustained outage (300/600 s), and full-run approach blackout. Timelines illustrate incoming lanes only. Purple marks and crosses indicate unavailable observations; hatched timeline entries are missing. Physical lanes remain open.}

\label{fig:7}

\end{figure*}

\subsection{5.3. Compared reconstruction and controller input methods}\label{compared-reconstruction-and-controller-input-methods}

Table V groups the compared inputs by role. Complete observation is the controller reference. Road Mean, Zero Fill, and Last Observed are deterministic point estimates. Predictive Mean, Fixed Lane Discount, and Staleness Gate share one frozen A2 checkpoint. The BRITS comparison uses a forward-only adaptation with observations available through the current decision. The MagiNet comparison uses a trailing 12-decision window ending at the current decision. A0 through A3 form the capacity-matched architecture comparison.

\begin{table*}[p]
\centering
\caption*{\textbf{Table V.} Compared controller inputs and architecture variants}
\label{tab:V}

\begin{tblr}{width=\linewidth,colspec={Q[0.1500,l,m] Q[0.1400,l,m] Q[0.2300,l,m] Q[0.0950,c,m] Q[0.2050,l,m] Q[0.1800,l,m]},
colsep=3.4pt,rowsep=3pt,cells={font=\TableBodyFont,valign=m},
rows={ht=32pt},row{1}={font=\TableBodyFont\bfseries,halign=c,ht=25pt},
hline{1,Z}={0.6pt},hline{2}={0.35pt},
row{2}={font=\TableBodyFont\bfseries\itshape,ht=19pt,rowsep=3pt},
hline{2,3}={0.25pt},
row{4}={font=\TableBodyFont\bfseries\itshape,ht=19pt,rowsep=3pt},
hline{4,5}={0.25pt},
row{8}={font=\TableBodyFont\bfseries\itshape,ht=19pt,rowsep=3pt},
hline{8,9}={0.25pt},
row{12}={font=\TableBodyFont\bfseries\itshape,ht=19pt,rowsep=3pt},
hline{12,13}={0.25pt},
row{15}={font=\TableBodyFont\bfseries\itshape,ht=19pt,rowsep=3pt},
hline{15,16}={0.25pt}}
Method & Category & Missing-lane value & Uncertainty & Decision-time information & Scope \\
\SetCell[c=6]{l} \emph{\textbf{Complete observation reference}} &  &  &  &  &  \\
Complete observation Signed Max-Pressure & Reference & Measured current lane count & No & Complete current observations & Clean matched reference \\
\SetCell[c=6]{l} \emph{\textbf{Deterministic controller inputs}} &  &  &  &  &  \\
Road Mean & Deterministic & Same-road observed mean; last-observed fallback & No & Same-road observations and stored lane history & Primary four-method matrix \\
Zero Fill & Deterministic & Zero & No & Current mask & Additional deterministic-baseline comparison \\
Last Observed & Deterministic & Last observed count; zero before first observation & No & Past observations and current mask & Additional deterministic-baseline comparison \\
\SetCell[c=6]{l} \emph{\textbf{LMP-GNN controller input rules}} &  &  &  &  &  \\
Predictive Mean & Probabilistic & Predictive mean (Eq. 13) & Marginal SD & Decision-time features and roadnet relations & Main severity and demand-variation studies \\
Fixed Lane Discount & Probabilistic & Mean minus one SD, clipped at zero (Eq. 14) & Used & Same A2 output and decision-time information & Main severity and demand-variation studies \\
Staleness Gate & Probabilistic & Staleness-weighted discount (Eq. 15) & Used & Same information plus decisions since last observation & Main severity and demand-variation studies \\
\SetCell[c=6]{l} \emph{\textbf{External learned adaptations}} &  &  &  &  &  \\
Decision-time BRITS adaptation & Recurrent adaptation & Forward-only point estimate & No & Forward-only information through the current decision & Gudang, 90 cells; decision-time local adaptation \\
Decision-time MagiNet adaptation & Graph adaptation & Endpoint estimate; Last Observed during initial history & No & Trailing 12-decision window ending at the current decision & Gudang, 90 cells; decision-time adaptation \\
\SetCell[c=6]{l} \emph{\textbf{Capacity-matched architecture comparison}} &  &  &  &  &  \\
A0 & No-graph variant & Mean and marginal variance from a no-graph encoder & Yes & Lane features & Architecture comparison \\
A1 & Lane-only variant & Mean and marginal variance from lane-only messages & Yes & Lane messages & Architecture comparison \\
\textbf{A2} & \textbf{Selected lane-movement variant} & \textbf{Mean and marginal variance from lane and movement messages} & \textbf{Yes} & \textbf{Lane and movement relations} & \textbf{Retained LMP-GNN architecture} \\
A3 & Phase-path comparison & Mean and marginal variance with an added phase path & Yes & Added learned phase relations & Comparison variant; A2 remains the proposed model \\
\end{tblr}

\end{table*}

All matched closed-loop comparisons use the same observed values, masks, controller, demand, and horizon. When a lane count is hidden at decision \(t\), a method does not receive that true count or any measurement collected after \(t\); the hidden count is used afterward only to calculate evaluation error. A point estimate is one numerical replacement for a missing lane without a predictive uncertainty distribution. Predictive Mean sends A2\textquotesingle s mean directly to SMP, while Fixed Lane Discount and Staleness Gate modify that mean. The evaluation therefore records both the original prediction and the final value that the controller receives. Likelihood and coverage are reported only for methods with probabilistic outputs.

\subsection{5.4. Evaluation metrics}\label{evaluation-metrics}

The metrics follow the implemented sequence. Reconstruction metrics evaluate the estimated missing count. Controller-input metrics evaluate the replacement value after an input rule is applied. In a fixed-state comparison, the stored traffic snapshot is held constant while complete and reconstructed inputs are used to recalculate pressure scores and phase choice. Closed-loop metrics summarize the traffic accumulated after repeated decisions. These stages use different units of analysis and are reported separately.

Predictive-mean MAE and RMSE are defined in Eqs. (16) and (17):

\setcounter{equation}{15}
\begin{equation}
\text{MAE}_{\mu}=\frac{1}{N_{M}}\sum_{(t,i) \in M}^{}\left| \mu_{t,i}-q_{t,i} \right|
\label{eq:16}
\end{equation}

\setcounter{equation}{16}
\begin{equation}
\text{RMSE}_{\mu}=\sqrt{\frac{1}{N_{M}}\sum_{(t,i) \in M}^{}\left( \mu_{t,i}-q_{t,i} \right)^{2}}
\label{eq:17}
\end{equation}

Gaussian NLL and one- or two-standard-deviation coverage are defined in Eqs. (18) and (19). The evaluation NLL includes the Gaussian constant, while the training objective in Eq. (12) omits it:

\setcounter{equation}{17}
\begin{equation}
\begin{gathered}\text{NLL=}\frac{1}{2N_{M}}\sum_{(t,i) \in M}^{}\Bigl[ \log(2\pi)+2log\sigma_{t,i}\\{}+\left( \frac{q_{t,i}-\mu_{t,i}}{\sigma_{t,i}} \right)^{2} \Bigr]\end{gathered}
\label{eq:18}
\end{equation}

\setcounter{equation}{18}
\begin{equation}
\begin{gathered}\text{Cov}_{k}=\frac{1}{N_{M}}\sum_{(t,i) \in M}^{}1\left( \left| q_{t,i}-\mu_{t,i} \right| \leq k\sigma_{t,i} \right),\\ k \in \{ 1,2\}\end{gathered}
\label{eq:19}
\end{equation}

Controller-input MAE evaluates the post-rule value actually supplied to SMP:

\setcounter{equation}{19}
\begin{equation}
\text{MAE}_{x}^{(k)}=\frac{1}{N_{M}}\sum_{(t,i) \in M}^{}\left| x_{t,i}^{(k)}-q_{t,i} \right|
\label{eq:20}
\end{equation}

At one fixed traffic snapshot, SMP calculates a pressure score for every legal phase. Pressure-vector MAE compares each reconstructed-input score with the corresponding complete-observation score and averages the absolute differences. Phase agreement is the proportion of decisions that select a phase in the complete-observation highest-score set. Wrong-phase rate is the remaining proportion. For example, 80\% phase agreement corresponds to a 20\% wrong-phase rate. When several complete-observation phases tie for the highest score, selecting any of them counts as agreement. Signed pressure bias, pairwise phase-ranking accuracy, score correlation, normalized regret, and top-two margin distortion provide secondary controller diagnostics.

The primary CityFlow endpoint is accrued average travel time (ATT). It includes completed trip travel time and the elapsed time to the horizon for generated vehicles that remain unfinished:

\setcounter{equation}{20}
\begin{equation}
\text{ATT}_{H}=\frac{\sum_{v \in C}^{}T_{v}+\sum_{v \in U}^{}\left( H-a_{v} \right)}{|C|+|U|}
\label{eq:21}
\end{equation}

Secondary CityFlow outcomes include waiting load and the number of admitted vehicles still inside the simulated road network at the 3,600 s horizon. This final vehicle count indicates uncleared traffic, excludes vehicles waiting in entry buffers, and is not a direct throughput measure. SUMO uses completed-trip ATT, detected load, and completion rate on its own scale. Run-level ATT medians and nearest-rank P90 summarize variation across simulator seeds. P90 describes the upper end of ATT across runs; it is not the 90th percentile of individual vehicle travel times.

\subsection{5.5. Experimental protocols and statistical analysis}\label{experimental-protocols-and-statistical-analysis}

The three analyses repeat different experimental conditions. The fixed-demand analysis treats each matched outage schedule as one observation. The demand-variation and geometry analyses treat each generated demand case as one observation. Confidence intervals and tests are calculated within these designs rather than by combining all simulation rows. Holm adjustment is applied within each predefined comparison family.

\subsubsection{5.5.1. Fixed-demand severity study}\label{fixed-demand-severity-study}

For each outage schedule, the candidate and comparator use the same network, demand, and missing observations. Their difference forms one paired observation. The paired \(t\) analysis summarizes the mean of the ten schedule-level differences, while the exact Wilcoxon analysis provides a rank-based sensitivity check. Equal-network summaries describe the five evaluated benchmarks.

\subsubsection{5.5.2. Evaluation under demand variation}\label{evaluation-under-demand-variation}

The demand-variation study asks whether two selected result directions persist when demand changes. It uses Synthetic 4 × 4, Arterial 1 × 6, and Gudang. For each generated demand case, two outage schedules are applied to both the candidate and comparator. The two candidate-minus-comparator ATT differences are averaged to give one demand-level observation. Repeating this process for 12 generated demands gives 12 observations for inference. Each demand is generated around a fixed one-hour base pattern that retains the original route and timing structure.

\subsubsection{5.5.3. Controlled comparison of missingness geometry}\label{controlled-comparison-of-missingness-geometry}

The geometry comparison asks whether grouping unavailable lanes into complete approaches changes relative method performance. Both mask constructions use the same eligible movement-start lanes, generated demand case, and number of missing lanes at every decision. The clustered mask hides complete approach groups, while the dispersed mask hides the same number of individual lanes across that eligible set. G1 compares Fixed Lane Discount with Predictive Mean, and G2 compares Staleness Gate with Road Mean. Eqs. (22) and (23) define the two geometry contrasts:

\setcounter{equation}{21}
\begin{equation}
G_{1}=\left( \text{ATT}_{\text{FLD}}^{C}-\text{ATT}_{\text{PM}}^{C} \right)-\left( \text{ATT}_{\text{FLD}}^{R}-\text{ATT}_{\text{PM}}^{R} \right)
\label{eq:22}
\end{equation}

\setcounter{equation}{22}
\begin{equation}
G_{2}=\left( \text{ATT}_{\text{SG}}^{C}-\text{ATT}_{\text{RM}}^{C} \right)-\left( \text{ATT}_{\text{SG}}^{R}-\text{ATT}_{\text{RM}}^{R} \right)
\label{eq:23}
\end{equation}

For each generated demand case, the mask replicates are averaged first. We calculate the candidate-minus-comparator ATT difference under clustered masks and under dispersed masks, then subtract the dispersed difference from the clustered difference. This difference of differences is the geometry contrast. A negative value means that the candidate gains more under clustered missingness. The 12 demand-level contrasts, rather than the underlying simulation rows, form the observations used in the statistical analysis.

The accompanying reproducibility files link each reported result to its network, demand case, mask schedule, model checkpoint, simulator setting, and metric definition.

\section{6. Results}\label{results}

Results are organized at three evaluation levels. Section 6.1 examines whether A2 reconstructs unavailable lane counts accurately and reports useful uncertainty. Section 6.2 asks whether the resulting inputs preserve SMP scores and phase choice at decision-time traffic states. Section 6.3 evaluates the traffic produced when the full decision cycle is repeated.

\subsection{6.1. Reconstruction performance}\label{reconstruction-performance}

\subsubsection{6.1.1. Lane-count accuracy}\label{lane-count-accuracy}

The evaluation starts from a stored traffic snapshot in which every lane count is known. Selected counts are hidden from A2 but retained separately as ground truth. The simulator is not advanced, so queues, legal phases, and traffic conditions remain unchanged while reconstruction accuracy is measured. This prevents earlier phase choices from creating different traffic states before the predictions are compared.

Across the broad fixed-snapshot evaluation, A2 achieved a predictive-mean MAE of 0.7873 vehicles and an RMSE of 1.1167 vehicles (Figure 8a). The MAE means that the reconstructed counts differed from the hidden values by about 0.79 vehicles on average. The larger RMSE reflects occasional larger deviations.

The seven traces include training, validation, and structurally unseen network roles. This broad evaluation establishes A2\textquotesingle s absolute reconstruction accuracy across 4,333,392 masked lane events in 372 trace, outage, and schedule cases. The aligned Gudang comparison then compares methods on the same hidden endpoints.

In the single-trajectory Gudang comparison, Last Observed produced the lowest MAE at 0.4241 vehicles, while A2 produced the lowest RMSE at 0.8114 vehicles and supplied lane-level uncertainty. Last Observed performed best on typical absolute error in this narrow test, whereas A2 reduced the influence of larger errors and provided probabilistic information.

\subsubsection{6.1.2. Uncertainty calibration}\label{uncertainty-calibration}

A2\textquotesingle s predicted uncertainty was informative and close to the Gaussian reference. Gaussian NLL evaluates the predicted count and uncertainty together, while coverage records how often the hidden count falls within one or two predicted standard deviations.

The constant-inclusive NLL was 1.3661 nats per event. Coverage was 73.20\% within one standard deviation, slightly above the 68.27\% Gaussian reference, so the central interval was somewhat conservative. Coverage within two standard deviations was 94.56\%, close to the 95.45\% reference (Figure 8b).

This calibration supports using the predicted standard deviation in Fixed Lane Discount and Staleness Gate. Deterministic inputs and the decision-time BRITS and MagiNet adaptations provide point estimates in the implemented comparisons, so they are evaluated through point error and downstream controller measures.

\begin{figure*}[!t]

\centering

\includegraphics[width=1.0\textwidth]{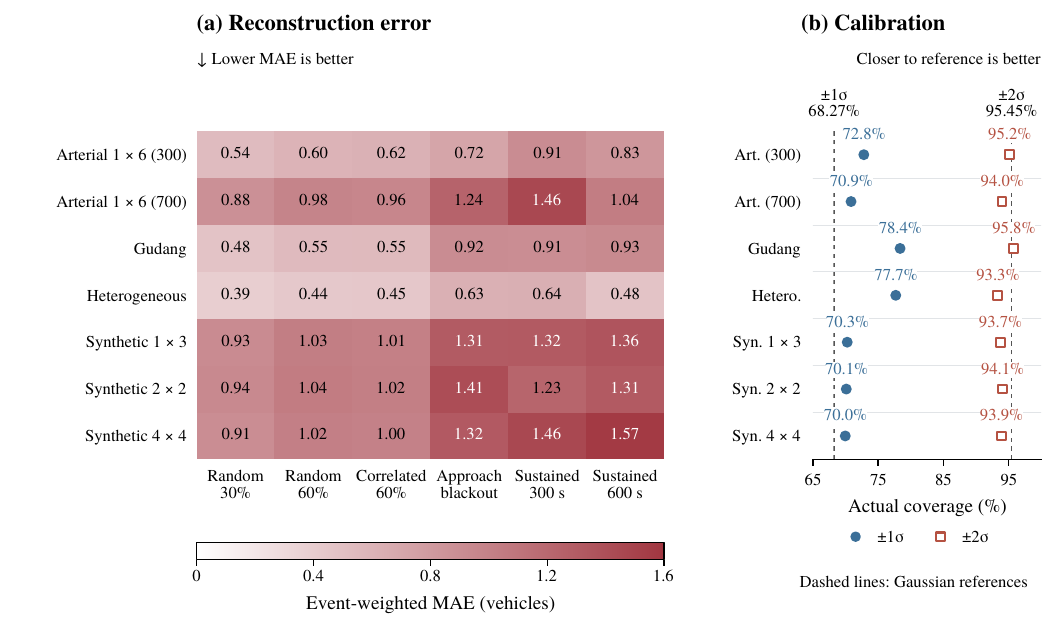}

\caption{A2 reconstruction and calibration across seven stored simulation traces. (a) Event-weighted MAE under six outage conditions; darker cells indicate larger errors. (b) Coverage within \ensuremath{\mu} ± \ensuremath{\sigma} and \ensuremath{\mu} ± 2\ensuremath{\sigma}, pooled across conditions with event weights; dashed lines are Gaussian references. Overall coverage is 73.20\% and 94.56\%, respectively.}

\label{fig:8}

\end{figure*}

\subsubsection{6.1.3. Prediction error and supplied-input error}\label{prediction-error-and-supplied-input-error}

Predictive-mean error measures how accurately A2 estimates the hidden count. Supplied-input error measures how close the final value received by SMP remains after an uncertainty rule modifies that estimate. Predictive Mean sends the model mean directly, while Fixed Lane Discount and Staleness Gate transform it before phase scoring.

Section 6.2 therefore evaluates the pressure scores and phase choice produced by the values that the controller actually receives.

\subsection{6.2. Pressure-score accuracy and phase agreement}\label{pressure-score-accuracy-and-phase-agreement}

At each decision along each method\textquotesingle s own closed-loop trajectory, SMP is evaluated twice at the same traffic state: once with the complete lane counts and once with the reconstructed input. The analysis compares the two legal-phase score vectors and then checks whether the reconstructed input selects one of the phases that has the highest score under complete observation.

Complete lane counts produce one pressure score for every legal phase. Every phase tied for the highest score belongs to the complete-observation highest-score set. Selecting any phase in this set counts as agreement. Wrong-phase rate is the proportion of evaluated decisions that select a phase outside that set. Lower values are better for both pressure-vector MAE and wrong-phase rate.

\subsubsection{6.2.1. How lane errors affect Signed Max-Pressure}\label{how-lane-errors-affect-signed-max-pressure}

A2-based inputs generally preserved pressure scores and phase choices better than Road Mean under low, moderate, and correlated high missingness (Figure 9). Predictive Mean produced the smallest pressure error through low and moderate severity, while Staleness Gate became strongest at the highest severities.

At random 60\%, Predictive Mean and Staleness Gate produced wrong-phase rates near 40.7\%, compared with 61.17\% for Road Mean. At random 90\%, Road Mean reached 70.77\%, slightly below the 72.05\% rate for Staleness Gate. The phase-agreement ranking therefore changed only in the most severe random condition.

\begin{figure*}[!t]

\centering

\includegraphics[width=0.94\textwidth]{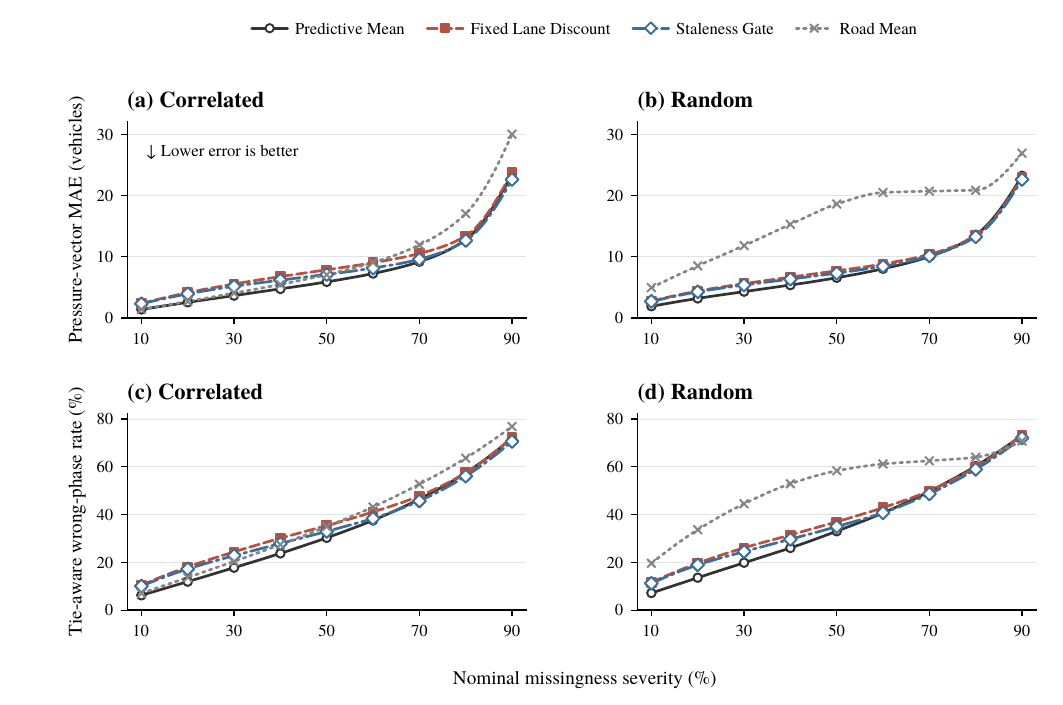}

\caption{Pressure-vector MAE (a, b) and tie-aware wrong-phase rate (c, d), comparing complete and reconstructed counts at the same state within each method's own closed-loop run. Values average ten masks per network, then five networks equally. Smooth PCHIP curves guide the eye between tested severities, without adding observations.}

\label{fig:9}

\end{figure*}

Lane-count MAE was a weak indicator of controller distortion across the 372 fixed-state cases. Pressure-vector MAE, in contrast, had a strong association with wrong-phase rate (Spearman rho = 0.929). Conditions with larger pressure distortion therefore tended to produce more phase disagreement.

The sampled decisions explain this difference. A small lane error can change the selected phase when two phase scores are close. A larger error can leave the winner unchanged when the leading phase has a wide margin, when signed errors cancel, or when the affected lane contributes little to the decisive comparison. A2-based inputs therefore preserved immediate SMP behavior better than Road Mean in most tested conditions, and pressure-score accuracy provided the clearest bridge from lane reconstruction to phase choice.

\subsubsection{6.2.2. Phase agreement and closed-loop traffic}\label{phase-agreement-and-closed-loop-traffic}

Phase agreement verifies whether reconstructed inputs preserve a complete-observation-optimal phase at the current decision-time traffic state. Closed-loop ATT then measures how those repeated choices accumulate through changing queues, observations, and later decisions. Across the matched conditions, better decision-time phase agreement did not consistently coincide with lower final accrued ATT. The near-zero pooled associations show that phase agreement is an important immediate controller check, while closed-loop evaluation measures the traffic created over the full horizon.

\subsection{6.3. Closed-loop traffic performance}\label{closed-loop-traffic-performance}

Closed-loop control repeats observation, reconstruction, pressure calculation, phase choice, and traffic evolution every 10 s. The primary CityFlow endpoint is accrued average travel time. It includes completed-trip travel time and the elapsed time to the 3,600 s horizon for generated vehicles that remain unfinished. Lower ATT means lower accumulated travel burden.

In Figure 10, ATT effects are reported relative to Road Mean. Negative values favor the candidate, while positive values favor Road Mean.

\subsubsection{6.3.1. Accrued ATT across missingness severity}\label{accrued-att-across-missingness-severity}

The strongest closed-loop benefit occurred under correlated missingness. Fixed Lane Discount reduced equal-network ATT by 1.152\% at correlated 10\%, 5.136\% at 60\%, and 13.743\% at 90\% (Figure 10). Predictive Mean and Staleness Gate were also favorable at high correlated severity, but Fixed Lane Discount produced the largest reduction.

Fixed Lane Discount also improved ATT under moderate random missingness, reaching 6.195\% lower ATT at random 60\%. The relative effects changed to +4.205\% at 80\% and +16.022\% at 90\%, so the ranking reversed at the two most severe random levels. Uncertainty-adjusted reconstruction therefore improved traffic under structured loss and moderate random loss.

\begin{figure*}[!t]

\centering

\includegraphics[width=0.97\textwidth]{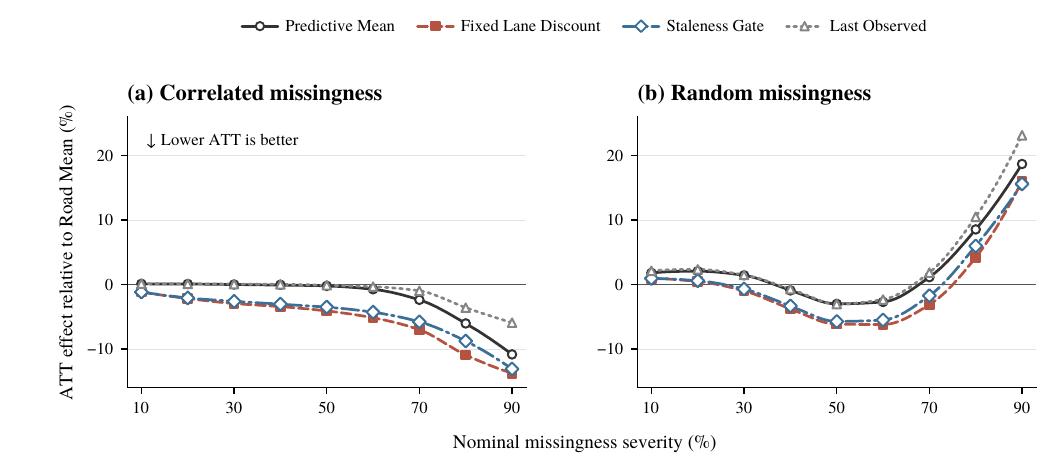}

\caption{CityFlow ATT effects relative to Road Mean; negative values favor the candidate. Paired percentage differences are averaged over masks within each network, then equally across five networks. Random and correlated masks have different eligible lane sets. Smooth PCHIP curves guide the eye between tested severities, without adding observations.}

\label{fig:10}

\end{figure*}

Correlated and random missingness differ in both spatial grouping and the lanes eligible to be hidden. Section 6.3.4 examines grouping directly by comparing masks that use the same eligible lane set and the same number of unavailable lanes.

\subsubsection{6.3.2. Network differences and secondary outcomes}\label{network-differences-and-secondary-outcomes}

The five-network average hides substantial network variation. At random 60\%, Staleness Gate lowered ATT on Synthetic 2 × 2, Synthetic 4 × 4, and Arterial 1 × 6, but raised ATT on Gudang and Heterogeneous. The effects ranged from -15.09\% to +4.24\% (Table VI), so the macro average does not identify the preferred input for every network.

\begin{table*}[!t]
\centering
\caption*{\textbf{Table VI.} Network-specific Staleness Gate effects at random 60\% relative to Road Mean}
\label{tab:VI}

\begin{tblr}{width=\linewidth,colspec={Q[0.2400,l,m] Q[0.1400,r,m] Q[0.1400,r,m] Q[0.2500,c,m] Q[0.2300,l,m]},
colsep=3.4pt,rowsep=3pt,cells={font=\TableBodyFont,valign=m},
rows={ht=25pt},row{1}={font=\TableBodyFont\bfseries,halign=c,ht=25pt},
hline{1,Z}={0.6pt},hline{2}={0.35pt}}
Fixed network & ATT effect (\%) & Difference (\(s\)) & 95\% interval (\(s\)) & Direction \\
Synthetic 2×2 & \textbf{-15.09\%} & \textbf{-22.14} & \textbf{{[}-24.70, -19.57{]}} & \textbf{Staleness Gate faster} \\
Synthetic 4×4 & \textbf{-14.28\%} & \textbf{-35.00} & \textbf{{[}-36.54, -33.47{]}} & \textbf{Staleness Gate faster} \\
Arterial 1×6 & \textbf{-3.51\%} & \textbf{-3.91} & \textbf{{[}-4.52, -3.30{]}} & \textbf{Staleness Gate faster} \\
Gudang & \ul{+4.24\%} & \ul{+13.98} & \ul{{[}+13.16, +14.80{]}} & \ul{Road Mean faster} \\
Heterogeneous & \ul{+1.20\%} & \ul{+3.53} & \ul{{[}+2.40, +4.66{]}} & \ul{Road Mean faster} \\
\end{tblr}

\vspace{4pt}
\begin{minipage}{\textwidth}\fontsize{8.5}{10}\selectfont \emph{Note. Negative values indicate lower Staleness Gate ATT. Bold entries indicate lower Staleness Gate ATT. Underlined entries indicate lower Road Mean ATT.}\end{minipage}
\end{table*}

ATT and waiting load describe different properties of the same traffic trajectory. At random 60\%, Predictive Mean lowered the equal-network ATT effect by 2.69\%, while waiting load increased by 19.71\%. ATT remains the primary travel-burden measure, and the load metrics provide additional congestion context.

The next two analyses vary demand and then compare clustered and dispersed missingness under a matched lane set. They test whether the selected traffic directions persist and how spatial arrangement changes the input comparison.

\subsubsection{6.3.3. Results under demand variation}\label{results-under-demand-variation}

The selected correlated-60 and random-90 directions remained unchanged across all 12 generated demand cases on all three tested networks. The two conditions were chosen before the new outcomes were examined.

For each generated demand case, two matched outage schedules were applied to both the candidate and comparator, and their two ATT differences were averaged to form one demand-level observation. This produced 12 observations for each network and contrast, with Holm adjustment across the 12 predefined tests.

Every comparison followed the expected direction, and every unadjusted 95\% interval excluded zero; the paired tests also remained significant after Holm adjustment (Figure 11). Under correlated 60\%, candidate inputs were 3.28 to 9.94 s faster than their comparators. Under random 90\%, Fixed Lane Discount and Staleness Gate were 26.94 to 68.60 s slower than Road Mean.

\begin{figure*}[!t]

\centering

\includegraphics[width=1.0\textwidth]{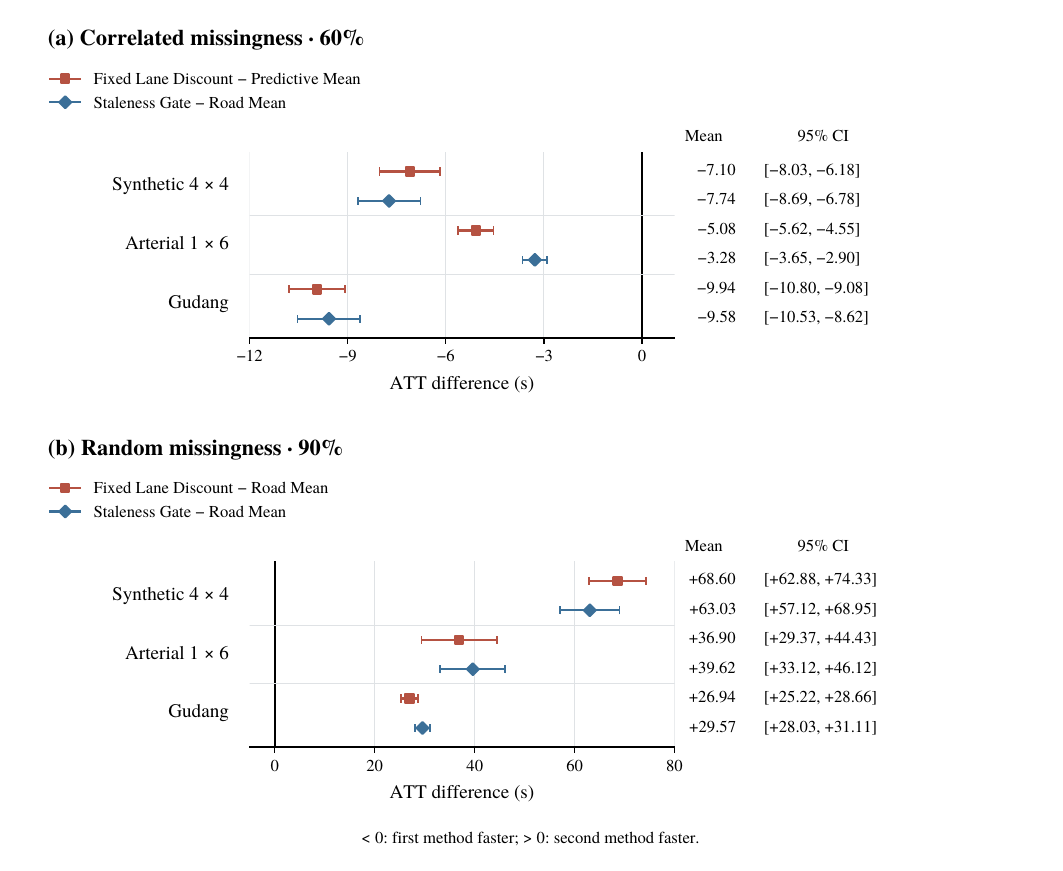}

\caption{ATT differences at correlated 60\% and random 90\% missingness. Points show means; bars show unadjusted paired 95\% t confidence intervals across 12 demand cases. Comparators and horizontal scales differ between panels.}

\label{fig:11}

\end{figure*}

These results show that the two anchor findings did not depend only on one fixed demand file. The remaining severity levels retain their fixed-demand results.

\subsubsection{6.3.4. Effects of missingness geometry}\label{effects-of-missingness-geometry}

Both mask constructions hide the same number of lanes from the same eligible movement-start-lane set under the same generated demand. The clustered mask hides complete approach groups, while the dispersed mask spreads the same missing count across individual lanes.

For each mask geometry, the method effect is candidate ATT minus comparator ATT. The geometry contrast is the clustered method effect minus the dispersed method effect. A negative contrast means that the candidate gains more under clustered missingness.

G1 compares Fixed Lane Discount with Predictive Mean. Its contrasts were negative on all three networks (-1.71, -0.96, and -2.31 s), showing a larger relative gain for Fixed Lane Discount when related lanes disappeared together. G2 compares Staleness Gate with Road Mean. Its contrast was positive on Synthetic 4 × 4 and Arterial 1 × 6, but negative on Gudang (Figure 12).

\begin{figure*}[!t]

\centering

\includegraphics[width=0.95\textwidth]{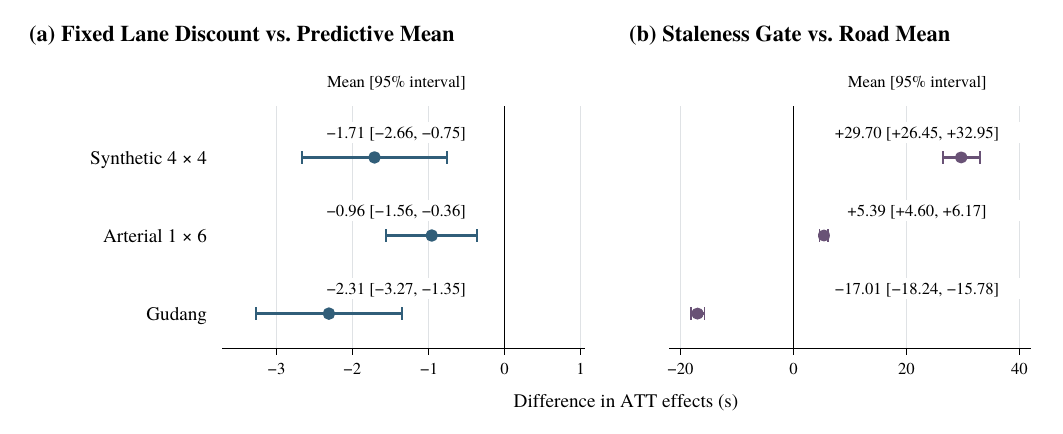}

\caption{Effects of missingness geometry. Points show clustered-minus-dispersed differences in candidate-minus-comparator ATT. Negative values relatively favor the candidate under clustered missingness; positive values under dispersed missingness. Bars: paired 95\% t intervals over 12 demand realizations. Panel scales differ.}

\label{fig:12}

\end{figure*}

The geometry result therefore shows that Fixed Lane Discount is especially effective relative to Predictive Mean when related lanes disappear together. The value of Staleness Gate also depends on the network and the available lane history, so its comparison with Road Mean changes direction.

\subsection{6.4. Additional comparisons}\label{additional-comparisons}

\subsubsection{6.4.1. Comparison with learned models}\label{comparison-with-learned-models}

Figure 13 reports learned-model ATT minus Staleness Gate ATT. Positive values favor Staleness Gate, while negative values favor the learned adaptation.

Staleness Gate outperformed both decision-time learned adaptations at random and correlated 20\% and 40\% missingness on Gudang. It remained faster than the decision-time BRITS adaptation at 60\%. At 80\%, the BRITS intervals included zero, and the approach-blackout interval also crossed zero.

The decision-time MagiNet adaptation became faster than Staleness Gate at 60\% and 80\%. Its approach-blackout interval crossed zero (Figure 13). The result is therefore a severity-dependent crossover on Gudang rather than an overall learned-model ranking.

\begin{figure*}[!t]

\centering

\includegraphics[width=0.95\textwidth]{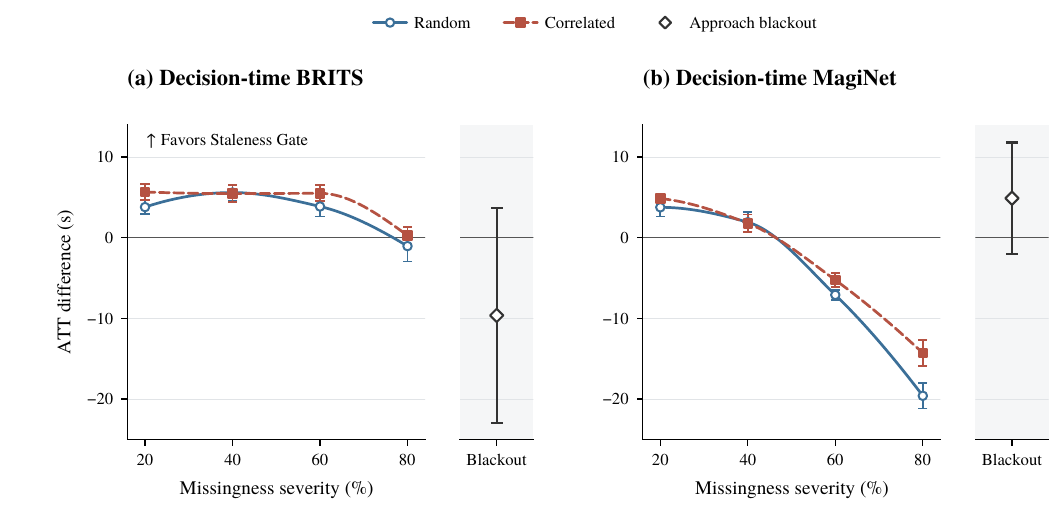}

\caption{Gudang adaptation ATT minus Staleness Gate ATT; positive values favor Staleness Gate. Points and bars show paired means and unadjusted 95\% t intervals: 10 mask schedules per severity, 9 for blackout. Smooth PCHIP curves guide the eye between tested severities, without adding observations. Blackout remains separate; both intervals cross zero, identifying no clearly faster method.}

\label{fig:13}

\end{figure*}

The crossover also shows why closed-loop traffic must be evaluated directly. Point-accuracy rankings and ATT rankings can differ after the reconstructed inputs pass through repeated control decisions.

\subsubsection{6.4.2. Architecture comparison and selection}\label{architecture-comparison-and-selection}

Lane-movement messages provided the strongest architecture improvement, while adding a learned phase path did not improve on A2. Relative to A0, A2 reduced masked MAE by 9.3\%, Gaussian NLL by 8.5\%, pressure loss by 24.8\%, ranking loss by 16.0\%, and first-max wrong-phase rate by 6.0\% (Figure 14a). A2 also improved every displayed measure over A1.

A3 was slightly worse than A2 on every displayed offline measure. Their closed-loop ATT-to-clean ratios differed by about 0.051\%: A2 reached 1.018382 and had lower mean ATT in 23 of 45 cells, while A3 reached 1.018903 and was lower in 22 cells. The additional phase-message path provided no consistent gain, so A2 remained the selected architecture.

\subsubsection{6.4.3. Model size and CPU inference latency}\label{model-size-and-cpu-inference-latency}

A2 was the smallest and fastest of the three models compared in Figure 14b--c. It contains 63,362 parameters, stores a 0.253 MiB checkpoint, and achieves 0.983 ms median one-thread CPU inference latency. The decision-time BRITS and MagiNet adaptations use 9.43 and 29.81 times as many parameters, with 5.28 and 19.98 times the median latency.

\begin{figure*}[!t]

\centering

\includegraphics[width=1.0\textwidth]{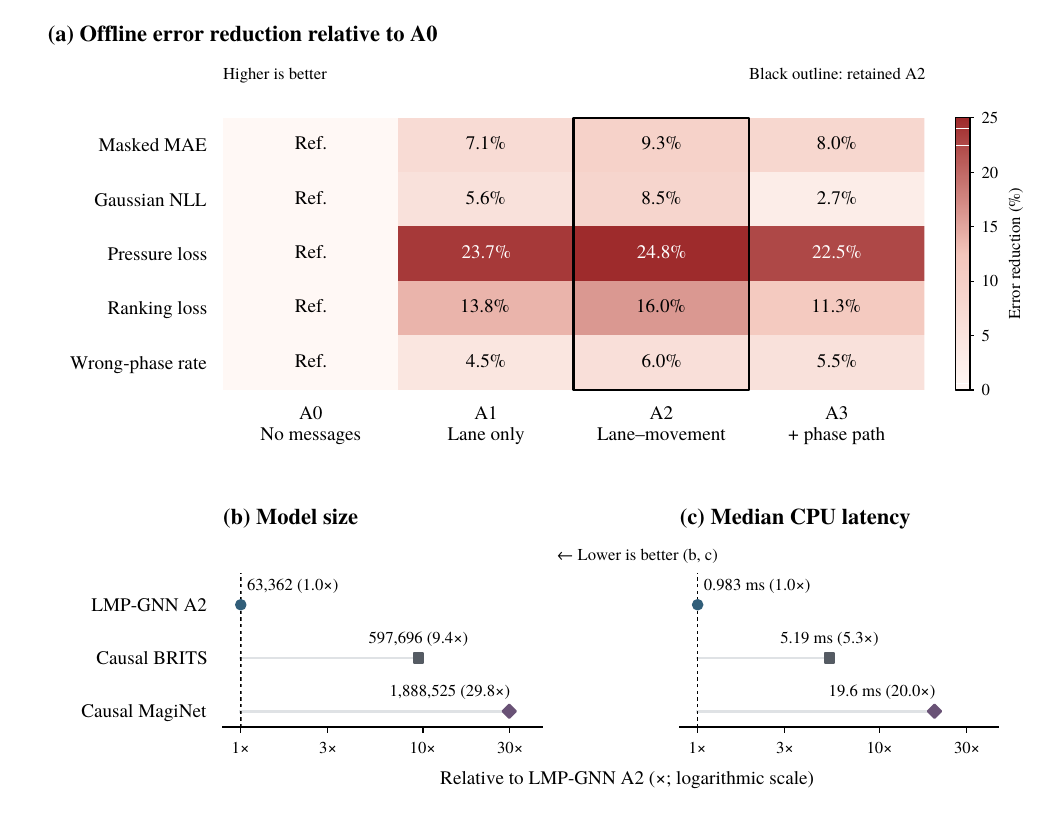}

\caption{Architecture error reductions relative to A0 (a), and parameter/median CPU-latency costs relative to A2 on log scales (b, c). Larger reductions and lower costs are better. A2 is retained. Panel a uses first-max wrong-phase rate and constant-inclusive Gaussian NLL.}

\label{fig:14}

\end{figure*}

All measured inference times are far below the 10 s control interval. A2 therefore leaves the most computational headroom for observation handling, feature construction, pressure calculation, communication, logging, and monitoring. The timing covers model inference and excludes feature and graph construction.

\subsubsection{6.4.4. Cross-simulator results}\label{cross-simulator-results}

The SUMO analysis compares each method with Road Mean within the same simulator because CityFlow and SUMO use different native ATT definitions. Raw ATT values are therefore kept separate.

The cross-simulator result was partially consistent. The favorable correlated-60 direction appeared in both CityFlow and SUMO, while random 60\% and approach blackout changed the Staleness Gate ranking between simulators.

Completion showed the same pattern. Correlated effects were nonnegative in both simulators, while random and blackout directions differed (Figure 15).

\begin{figure*}[!t]

\centering

\includegraphics[width=1.0\textwidth]{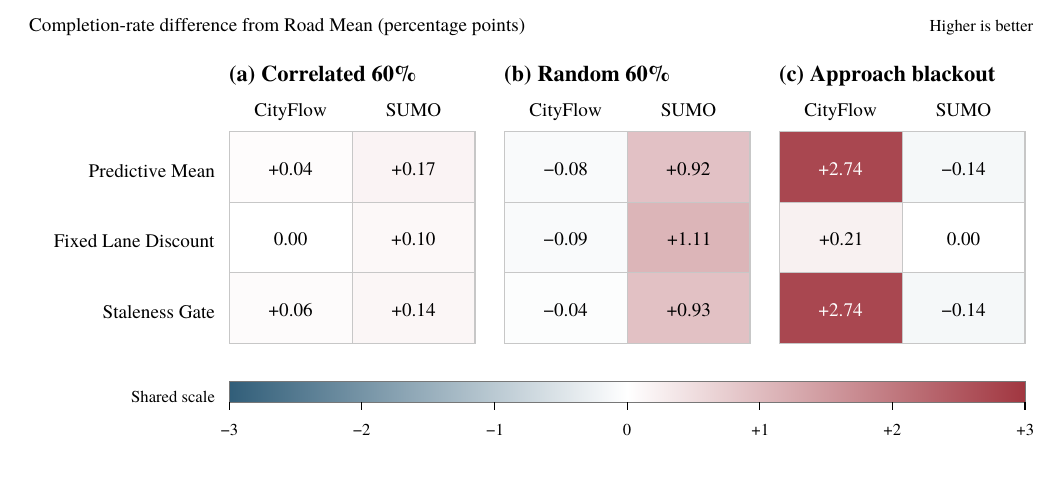}

\caption{Completion effects relative to Road Mean. Cells are descriptive candidate-minus-Road Mean differences in percentage points, averaged over ten within-simulator schedules. CityFlow and SUMO remain separate; all panels share a zero-centered color scale. Positive values favor the candidate.}

\label{fig:15}

\end{figure*}

The correlated result was therefore reproduced under both simulator implementations. The dispersed random and persistent blackout conditions remained simulator-dependent, which makes local calibration and validation important before transfer to a new environment.

\subsection{6.5. Summary of main results}\label{summary-of-main-results}

Taken together, the results show that A2 provides an accurate and efficient probabilistic input for unavailable lanes. Its reconstructed values generally preserved SMP pressure scores and phase choices more closely than Road Mean, while the traffic benefit depended on the missingness regime. Fixed Lane Discount was strongest under correlated missingness and remained favorable at moderate random loss, whereas the ranking changed under severe random loss. On Gudang, Staleness Gate was competitive with the learned adaptations at low and moderate severity. The architecture and runtime audits also supported A2 as the retained model, and the favorable correlated direction appeared in both CityFlow and SUMO.

\section{7. Discussion}\label{discussion}

Accurate lane reconstruction becomes useful to traffic control when the reconstructed values preserve the signed phase-pressure calculation and support effective decisions over repeated control cycles. The following sections explain this reconstruction-to-control chain, the conditions that shape traffic benefits, the methodological and practical implications, and the remaining limitations.

\subsection{7.1. From reconstruction to control}\label{from-reconstruction-to-control}

The reconstruction audit established accurate lane recovery. The control consequence of an error, however, depends on where the affected lane enters the SMP calculation. Upstream and downstream errors enter pressure with opposite signs, one lane may appear in several laneLinks, and signed errors from different lanes may reinforce or cancel. The pressure margin between competing phases then determines whether the selected phase changes.

Pressure-score accuracy provides the clearest bridge from reconstructed lane values to phase choice. Complete lane counts produce one score for every legal phase, and every phase tied for the highest score belongs to the complete-observation highest-score set. The strong association between pressure-vector MAE and wrong-phase rate (Spearman rho = 0.929) shows that larger score distortion usually produces more phase disagreement across the evaluated conditions.

Phase agreement verifies whether reconstructed inputs preserve a complete-observation-optimal phase at the current decision-time traffic state. Closed-loop ATT then measures how those repeated choices accumulate through changing queues, observations, and later decisions. Read together, the two measures distinguish immediate consistency with the established controller from the traffic benefit created over the full horizon. Their near-zero pooled associations with final ATT show why both evaluations are needed.

Gudang shows that method rankings change as the evaluation moves from lane reconstruction to closed-loop traffic. Last Observed leads on MAE in the single-trajectory comparison, while A2 leads on RMSE and supplies lane-level uncertainty. Staleness Gate and the decision-time MagiNet adaptation exchange the ATT lead as missingness becomes more severe. Reconstruction metrics identify component quality, while pressure scores, phase agreement, and closed-loop traffic determine whether that quality becomes a control benefit.

\subsection{7.2. Traffic benefits across missingness regimes}\label{traffic-benefits-across-missingness-regimes}

The spatial pattern of missing observations, the traffic context that remains available, and the network structure jointly shape the traffic benefit of LMP-GNN. Under correlated missingness, complete approach groups become unavailable together, while lane-movement relations still connect the affected approach with related incoming and outgoing traffic. A2 uses this context to estimate the missing counts, and Fixed Lane Discount reduces the values supplied when those estimates are uncertain. The observed pattern indicates that lane-movement context and uncertainty adjustment are particularly useful when related lanes become unavailable together.

Under severe random missingness, unavailable observations are dispersed across more locations and coherent local context becomes sparse. Fixed Lane Discount remains favorable at random 60\%, but the ranking changes at 80\% and 90\% as the dispersed loss becomes extreme. The principal traffic benefit therefore appears under correlated loss and moderate random loss.

The favorable correlated-60 result and the adverse random-90 result reappeared as demand intensity and timing varied around the base demand pattern on all three tested networks. Their direction remained stable across the 12 generated demand cases rather than appearing only under one fixed flow realization. This additional evaluation covers the two selected anchor conditions.

The geometry comparison separates spatial arrangement from the number of missing lanes. When the eligible lanes and missing count were held constant, grouping the unavailable lanes into complete approaches consistently improved Fixed Lane Discount relative to Predictive Mean. Related reconstruction errors can enter several pressure terms together, while the uncertainty discount limits the values supplied before those terms are summed. The Staleness Gate comparison changed direction by network because its adjustment also depends on the available lane history and network-specific pressure context. Network-specific effects and the MagiNet crossover therefore show that the preferred input rule changes with the operating condition, motivating future monitoring and predetermined rule switching.

\subsection{7.3. Methodological and practical implications}\label{methodological-and-practical-implications}

The main methodological contribution is an attributable reconstruction-to-control interface. Every compared method is converted into the same completed lane-count input expected by SMP. Because the downstream decision rule is common to all methods, differences in pressure, phase choice, and traffic can be traced to how each method fills the unavailable entries.

The completed lane-count interface is also reusable. Valid measurements retain priority, while reconstructed entries use the same lane granularity as measured inputs. Future studies can therefore evaluate the reconstruction layer with controller variants that accept a compatible lane-count state. The current work establishes this comparison with ordinary SMP.

A2\textquotesingle s architecture and runtime results support decision-time use. Lane-movement messages provide the main architecture gain, while the added learned phase path provides no consistent improvement. The 63,362-parameter model and 0.983 ms median CPU inference latency leave room within the 10 s control interval for sensing, feature construction, pressure calculation, communication, logging, and monitoring.

Staleness and predictive uncertainty provide useful monitoring signals for a future deployment. Staleness is the number of signal decisions since a lane was last observed, while the predicted standard deviation quantifies uncertainty in the reconstructed count. These signals may support predetermined switching or fallback rules in future work. The current study does not implement that mechanism.

The SUMO results show that transfer is condition-dependent. The favorable correlated direction appears in both simulators, while random and blackout rankings change. A new simulator or field site therefore requires local calibration and validation.

\subsection{7.4. Limitations and future work}\label{limitations-and-future-work}

The traffic evidence is entirely simulation-based, and the study does not yet include a live intersection or an observed detector-failure experiment. The main severity study covers five fixed CityFlow networks with one demand file per network, while the demand-variation analysis changes traffic around one base pattern on three networks and retests two anchor regimes. Future work should use independent traffic days, real detector outages, additional routes and phase plans, more network structures, hardware-in-the-loop evaluation, and monitored field trials.

The reported intervals describe outage and simulation variation for one selected A2 checkpoint. They do not include variability caused by retraining the model from different initializations. More outage schedules, independent demand conditions, and repeated training seeds would provide a fuller account of model and traffic variability.

The current reconstruction rules are transparent but fixed. Fixed Lane Discount and Staleness Gate use common parameters across networks and missingness regimes, and the study does not learn an online policy for switching among the three rules. A2 reports marginal uncertainty for individual lanes without modeling cross-lane covariance or joint phase-level uncertainty, and its training data come mainly from synthetic simulation traces rather than observed detector failures. Keeping SMP as the common controller supports attribution, but the current experiments do not test whether joint controller adaptation could improve the most severe missingness conditions.

The learned-model and runtime conclusions follow the implemented comparison scope. The BRITS and MagiNet models are decision-time adaptations rather than exact reproductions of their original bidirectional or full-window systems. The study does not provide a new SMP stability theorem or a control-safety guarantee, and the latency benchmark covers model inference rather than the complete observation-to-decision pipeline. Future work should evaluate complete online implementations, decision-aware or joint uncertainty, full pipeline timing, monitored failover, and staged deployment.

\section{8. Conclusion}\label{conclusion}

Missing lane observations can distort SMP because the controller calculates phase pressures from incomplete lane counts and may select a different phase. This study developed LMP-GNN, a lightweight lane-movement probabilistic reconstruction model that estimates an unavailable lane count together with lane-level uncertainty. Predictive Mean uses the estimate directly, Fixed Lane Discount subtracts an uncertainty penalty, and Staleness Gate adjusts that penalty using the time since the last valid observation. The main innovation is a selective lane-level probabilistic reconstruction interface that uses lane-movement context, preserves measured counts, and connects reconstructed values to SMP decisions and closed-loop traffic.

A2 accurately recovered unavailable lane counts and produced informative lane-level uncertainty. Its input rules generally improved pressure-score accuracy and phase agreement relative to Road Mean. Fixed Lane Discount reduced accrued average travel time by up to 13.74\% under correlated missingness and remained favorable under moderate random loss, while the ranking changed at the most severe random levels. Staleness Gate was competitive with the decision-time learned adaptations at low and moderate severity. The retained A2 architecture was substantially smaller and faster than those adaptations, supporting decision-time use.

The study establishes a controller-level evaluation that connects reconstruction accuracy with pressure scores, phase agreement, and closed-loop traffic outcomes. Future work should extend the analysis to independent traffic demand, additional network structures, observed detector failures, and field evaluation. Decision-aware or joint uncertainty, adaptive rule selection, monitored fallback, complete observation-to-decision timing, and controller variants that accept compatible lane-count inputs provide the next steps toward practical deployment.

\balance\section*{References}\begin{enumerate}[label={[\arabic*]},leftmargin=1.65em,itemsep=3pt plus 4pt,parsep=0pt]\fontsize{8.5}{10.2}\selectfont\interlinepenalty=10000

\item U.S. Federal Highway Administration, Traffic Detector Handbook: Third Edition---Volume II, Report No. FHWA-HRT-06-139, October 2006. \url{https://www.fhwa.dot.gov/publications/research/operations/its/06139/chapt6.cfm}

\item T. Wongpiromsarn, T. Uthaicharoenpong, Y. Wang, E. Frazzoli, D. Wang, Distributed traffic signal control for maximum network throughput, in: 2012 15th International IEEE Conference on Intelligent Transportation Systems (ITSC), IEEE, 2012, pp. 588--595. \url{https://doi.org/10.1109/ITSC.2012.6338817}

\item P. Varaiya, Max pressure control of a network of signalized intersections, Transportation Research Part C: Emerging Technologies 36 (2013) 177--195. \url{https://doi.org/10.1016/j.trc.2013.08.014}

\item H. Mei, J. Li, B. Shi, H. Wei, Reinforcement learning approaches for traffic signal control under missing data, in: Proceedings of the Thirty-Second International Joint Conference on Artificial Intelligence (IJCAI), 2023, pp. 2261--2269. \url{https://doi.org/10.24963/ijcai.2023/251}

\item H. Chen, Y. Jiang, S. Guo, X. Mao, Y. Lin, H. Wan, DiffLight: A partial rewards conditioned diffusion model for traffic signal control with missing data, Advances in Neural Information Processing Systems 37 (2024) 123353--123378. \url{https://doi.org/10.52202/079017-3921}

\item D. Xu, Z. Yu, X. Liao, H. Guo, A graph deep reinforcement learning traffic signal control for multiple intersections considering missing data, IEEE Transactions on Vehicular Technology 73 (12) (2024) 18307--18319. \url{https://doi.org/10.1109/TVT.2024.3444475}

\item M. Li, J. Wang, G. Yu, X. Wang, Q. Chen, W. Ni, L. Li, H. Peng, RobustLight: Improving robustness via diffusion reinforcement learning for traffic signal control, Proceedings of Machine Learning Research 267 (2025) 36192--36214. \url{https://proceedings.mlr.press/v267/li25cs.html}

\item L. Li, V. Okoth, S.E. Jabari, Backpressure control with estimated queue lengths for urban network traffic, IET Intelligent Transport Systems 15 (2) (2021) 320--330. \url{https://doi.org/10.1049/itr2.12027}

\item C. Tan, D. Sun, H. Liu, M. Rinaldi, H. van Lint, CV-MP: Max-pressure control in heterogeneously distributed and partially connected vehicle environments, Transportation Research Part B: Methodological 204 (2026) 103387. \url{https://doi.org/10.1016/j.trb.2025.103387}

\item Y. Tashiro, J. Song, Y. Song, S. Ermon, CSDI: Conditional score-based diffusion models for probabilistic time series imputation, Advances in Neural Information Processing Systems 34 (2021) 24804--24816. \url{https://proceedings.neurips.cc/paper/2021/hash/cfe8504bda37b575c70ee1a8276f3486-Abstract.html}

\item Z. Wang, D. Zhuang, Y. Li, J. Zhao, P. Sun, S. Wang, Y. Hu, ST-GIN: An uncertainty quantification approach in traffic data imputation with spatio-temporal graph attention and bidirectional recurrent united neural networks, in: IEEE 26th International Conference on Intelligent Transportation Systems (ITSC), IEEE, 2023, pp. 1454--1459. \url{https://doi.org/10.1109/ITSC57777.2023.10422526}

\item J. Zhou, B. Lu, Z. Liu, S. Pan, X. Feng, H. Wei, G. Zheng, X. Wang, C. Zhou, MagiNet: Mask-aware graph imputation network for incomplete traffic data, ACM Transactions on Knowledge Discovery from Data 19 (7) (2025) 130:1--130:20. \url{https://doi.org/10.1145/3743141}

\item F. Rodrigues, C.M. Lima Azevedo, Towards robust deep reinforcement learning for traffic signal control: Demand surges, incidents and sensor failures, in: 2019 IEEE Intelligent Transportation Systems Conference (ITSC), IEEE, 2019, pp. 3559--3566. \url{https://doi.org/10.1109/ITSC.2019.8917451}

\item T. Shi, F.-X. Devailly, D. Larocque, L. Charlin, Improving the generalizability and robustness of large-scale traffic signal control, IEEE Open Journal of Intelligent Transportation Systems 5 (2024) 2--15. \url{https://doi.org/10.1109/OJITS.2023.3331689}

\item R. Zhang, A. Ishikawa, W. Wang, B. Striner, O.K. Tonguz, Using reinforcement learning with partial vehicle detection for intelligent traffic signal control, IEEE Transactions on Intelligent Transportation Systems 22 (1) (2021) 404--415. \url{https://doi.org/10.1109/TITS.2019.2958859}

\item Q. Jiang, M. Qin, H. Zhang, X. Zhang, W. Sun, BlindLight: High robustness reinforcement learning method to solve partially blinded traffic signal control problem, IEEE Transactions on Intelligent Transportation Systems 25 (11) (2024) 16625--16641. \url{https://doi.org/10.1109/TITS.2024.3416154}

\item W. Cao, D. Wang, J. Li, H. Zhou, L. Li, Y. Li, BRITS: Bidirectional recurrent imputation for time series, Advances in Neural Information Processing Systems 31 (2018) 6775--6785. \url{https://proceedings.neurips.cc/paper/2018/hash/734e6bfcd358e25ac1db0a4241b95651-Abstract.html}

\item A. Cini, I. Marisca, C. Alippi, Filling the gaps: Multivariate time series imputation by graph neural networks, in: International Conference on Learning Representations (ICLR), 2022. \url{https://openreview.net/forum?id=kOu3-S3wJ7}

\item M. Liu, H. Huang, H. Feng, L. Sun, B. Du, Y. Fu, PriSTI: A conditional diffusion framework for spatiotemporal imputation, in: 2023 IEEE 39th International Conference on Data Engineering (ICDE), IEEE, 2023, pp. 1927--1939. \url{https://doi.org/10.1109/ICDE55515.2023.00150}

\item I. Marisca, A. Cini, C. Alippi, Learning to reconstruct missing data from spatiotemporal graphs with sparse observations, Advances in Neural Information Processing Systems 35 (2022) 32069--32082. \url{https://doi.org/10.52202/068431-2324}

\item Y. Wang, H. Peng, S. Wang, H. Du, C. Liu, J. Wu, G. Wu, STAMImputer: Spatio-temporal attention MoE for traffic data imputation, in: Proceedings of the Thirty-Fourth International Joint Conference on Artificial Intelligence, 2025, pp. 3435--3443. \url{https://doi.org/10.24963/ijcai.2025/382}

\item H. Zhang, S. Feng, C. Liu, Y. Ding, Y. Zhu, Z. Zhou, W. Zhang, Y. Yu, H. Jin, Z. Li, CityFlow: A multi-agent reinforcement learning environment for large scale city traffic scenario, in: The World Wide Web Conference (WWW), ACM, 2019, pp. 3620--3624. \url{https://doi.org/10.1145/3308558.3314139}

\item P.A. Lopez, M. Behrisch, L. Bieker-Walz, J. Erdmann, Y.-P. Flötteröd, R. Hilbrich, L. Lücken, J. Rummel, P. Wagner, E. Wiessner, Microscopic traffic simulation using SUMO, in: 2018 21st International Conference on Intelligent Transportation Systems (ITSC), IEEE, 2018, pp. 2575--2582. \url{https://doi.org/10.1109/ITSC.2018.8569938}

\end{enumerate}

\end{document}